\documentclass[twocolumn]{aastex62}
\usepackage{natbib}
\usepackage{xspace}
\graphicspath{{./}{pictures/}}

\received{}
\revised{}
\accepted{}
\submitjournal{ApJ}

\shorttitle{Tidal disruption events}
\shortauthors{Ryu et al.}

\usepackage{graphicx}	
\usepackage{amsmath}	
\usepackage{amssymb}	
\usepackage{cases}
\usepackage{array,multirow}
\usepackage{upgreek}
\usepackage{textcomp}
\usepackage{hyperref}
\usepackage[referable]{threeparttablex}
\usepackage{booktabs, dcolumn}

\newcommand{\beq}{\begin{equation}}
\newcommand{\eeq}{\end{equation}}
\newcommand{\simlt}{\mathrel{\hbox{\rlap{\hbox{\lower4pt\hbox{$\sim$}}}\hbox{$<$}}}}
\newcommand{\simgt}{\mathrel{\hbox{\rlap{\hbox{\lower4pt\hbox{$\sim$}}}\hbox{$>$}}}}

\newcommand{\s}{\;\mathrm{s}}

\newcommand{\Msol}{\;\mathrm{M}_{\odot}}
\newcommand{\Rsol}{\;\mathrm{R}_{\odot}}
\newcommand{\gram}{\;\mathrm{g}}
\newcommand{\cm}{\;\mathrm{cm}}

\newcommand{\km}{\;\mathrm{km}}

\newcommand{\pc}{\;\mathrm{pc}}
\newcommand{\yr}{\;\mathrm{yr}}
\newcommand{\Myr}{\;\mathrm{Myr}}

\newcommand{\K}{\;\mathrm{K}}

\newcommand{\rtidal}{\;r_{\rm t}}

\newcommand{\physrad}{\mathcal{R}_{\rm t}}
\newcommand{\harm}{{\sc Harm3d}}   
\newcommand{\mesa}{{\small MESA}}

\def\apjl{ApJL}
\def\apj{ApJ}
\def\mnras{M.N.R.A.S.}
\def\aap{A\&A}
\def\nat{Nat.}

\def\araa{Ann. Rev. A\&A}

\def\apjs{ApJ Supp.}

\defcitealias{Ryu1+2019}{Paper 1}
\defcitealias{Ryu2+2019}{Paper 2}
\defcitealias{Ryu3+2019}{Paper 3}
\defcitealias{Ryu4+2019}{Paper 4}

\begin{document}

\title{Tidal disruptions of main sequence stars - III. Stellar mass dependence of the character of partial disruptions}

\correspondingauthor{Taeho Ryu}
\email{tryu2@jhu.edu}

\author[0000-0002-0786-7307]{Taeho Ryu}
\affil{Physics and Astronomy Department, Johns Hopkins University, Baltimore, MD 21218, USA}

\author{Julian Krolik}
\affiliation{Physics and Astronomy Department, Johns Hopkins University, Baltimore, MD 21218, USA}
\author{Tsvi Piran}
\affiliation{Racah Institute of Physics, Hebrew University, Jerusalem 91904, Israel}

\author{Scott C. Noble}
\affiliation{Gravitational Astrophysics Laboratory, Goddard Space Flight Center, Greenbelt, MD 20771, USA}

\begin{abstract}

In this paper, the third in this series, we continue our study of tidal disruption events of main-sequence stars by a non-spinning $10^{6}~\rm{M}_\odot$ supermassive black hole. 
Here we focus on the stellar mass dependence of the outcomes of partial disruptions. As the encounter becomes weaker, the debris mass is increasingly concentrated near the outer edges of the energy distribution. 
As a result, the mass fallback rate can deviate substantially from a $t^{-5/3}$ power-law, becoming more like a single peak with a tail declining as $t^{-p}$ with $p\simeq2-5$.  Surviving remnants are spun-up in the prograde direction and are hotter than main sequence stars of the same mass. Their specific orbital energy is $\simeq10^{-3}\times$ that of the debris, but of either sign with respect to the black hole potential, while their specific angular momentum is close to that of the original star. Even for strong encounters, remnants have speeds at infinity relative to the black hole potential $\lesssim 300$~km~s$^{-1}$, so they are unable to travel far out into the galactic bulge.  The remnants most deeply bound to the black hole go through a second tidal disruption event upon their first return to pericenter; if they have not thermally relaxed, they will be completely disrupted.
\end{abstract}


\section{Introduction} \label{sec:intro}

\begin{figure*}
	\centering
\includegraphics[width=14.0cm]{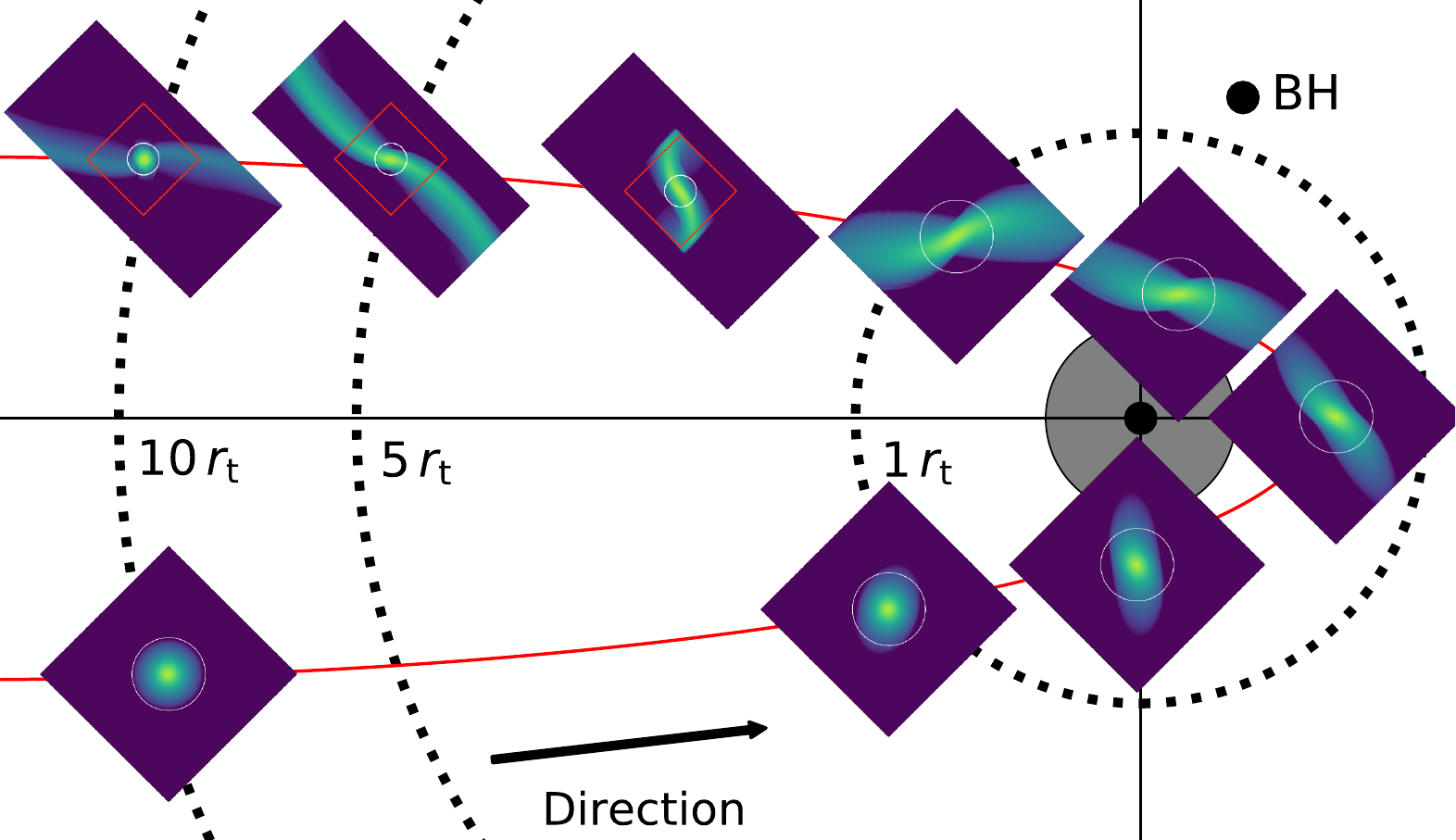}
\caption{Snapshots showing successive moments for a partial disruption ($M_{\star}=1$ and $r_{\rm p}=0.55\rtidal$). The red line indicates the star's orbit around the black hole (black circle) whose pericenter is larger than the physical tidal radius $\mathcal{R}_{\rm t}=0.475\rtidal$ (shaded circle). Each inset figure shows the stellar density distribution in the orbital plane and the shape of the simulation box. The initially cubic box is replaced by a rectangular box when the orbital separation $r>2-3\rtidal$ (See \citetalias{Ryu2+2019} for more details). The white circle in the center of each box depicts the initial stellar radius, and the red square in the rectangular boxes shows the position and size of the original cubic box. Note that the rectangular boxes are not drawn to scale with the cubic boxes; the dotted curves indicating $1\rtidal$, $5\rtidal$ and $10\rtidal$ are also not drawn to scale.}
\label{fig:overview}
\end{figure*}

Supermassive black holes (SMBHs) exert a significant tidal gravity on stars when their separation becomes comparable to or shorter than the ``tidal radius''.  Only if the star passes inside the physical tidal radius $\physrad$ is it fully disrupted; otherwise, if its pericenter $r_{\rm p}\gtrsim\physrad$, it is partially disrupted and loses only a fraction of its mass. In both cases, roughly half of the mass removed from the star is bound to the black hole. When the bound debris returns to the vicinity of the BH, it may produce a luminous flare. 

This is the third paper in a series of four whose aim is to study quantitatively the key properties of tidal disruption events (TDEs) as a function of stellar mass $M_{\star}$ and black hole mass $M_{\rm BH}$.  To do so, we have performed a suite of hydrodynamic simulations employing the intrinsically-conservative grid-based general relativistic hydrodynamics code \harm \citep{Noble+2009}.  With initial data for the stars created using main-sequence models generated by {\small MESA}, we compute the time-dependent stellar self-gravity in relativistically consistent fashion (further methodological details can be found in \citealt{Ryu2+2019}).   This apparatus is then applied to events involving stars of eight different masses, ranging from $0.15\Msol$ to $10\Msol$, and with multiple pericenter distances $r_{\rm p}$ for each stellar mass.

In this paper, we focus on
how the outcomes of partial disruptions (surviving remnants and stellar debris) depend on stellar mass $M_{\star}$ and orbital pericenter $r_{\rm p}$ when the black hole has no spin and mass $M_{\rm BH} = 10^6$ (from this point on, all masses will be given in solar mass units).
We provide a short overview of our simulation setup in Section~\ref{sec:sim_setup}.
In Section~\ref{sec:results}, we present the distribution of energy and the fallback rate of stellar debris (Section~\ref{subsub:dmde_partial}). Then we analyze the properties of the surviving remnants (Section~\ref{subsec:remnant}): the mass of surviving remnants for different degrees of partial disruption (Section~\ref{subsub:mass_remnant}); the specific orbital energy of the remnants (Section~\ref{subsub:bound_unboud}); remnant spin (Section \ref{res:internal}) and remnant internal structure (Section ~\ref{subsub:density}). We discuss the future fate of partially disrupted stars in Section~\ref{discuss:partialTD}. Finally, we conclude with a summary of our findings in Section~\ref{sec:summary}.

Throughout this paper, symbols with the subscript $\star$, such as $R_{\star}$ (stellar radius) and $M_{\star}$ (stellar mass), always refer to the properties of the star at the beginning of the tidal encounter. All masses are measured in units of ${\rm M}_\odot$ and stellar radii in units of ${\rm R}_\odot$.

\begin{figure*}
	\centering
	\includegraphics[width=8.9cm]{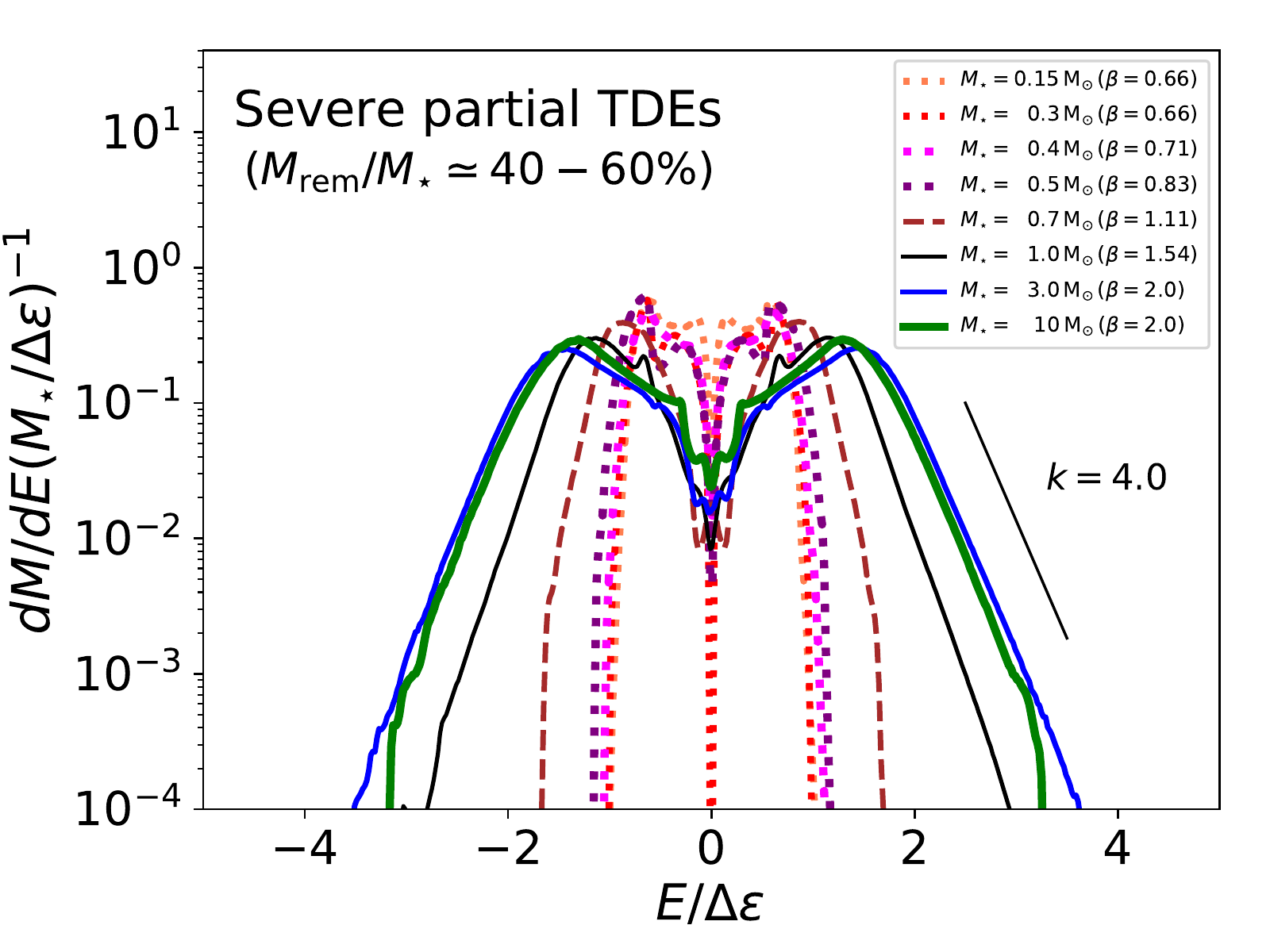}
	\includegraphics[width=8.9cm]{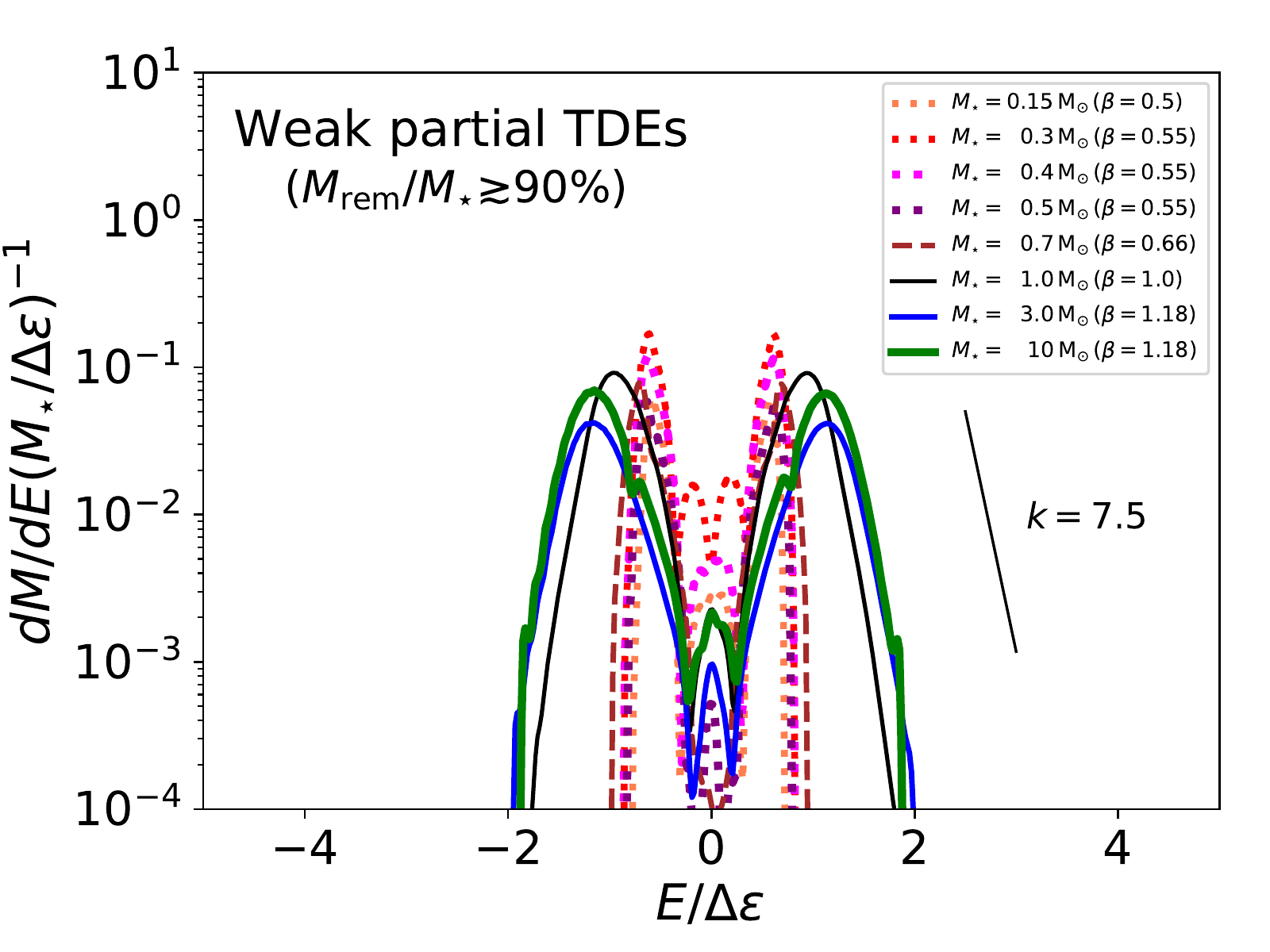}	
	\caption{$dM/dE$ for the stellar debris produced in partial TDEs with $M_{\rm rem}/M_{\star}\simeq40-60\%$ (\textit{left} panel) and $\gtrsim90\%$ (\textit{right} panel). We normalize the distribution with $M_{\star}/\Delta\epsilon$, where $\Delta\epsilon=G M_{\rm BH}R_{\star}/r_{\rm t}^{2}$. The integrated area under each curve is therefore the fractional mass of the stellar debris $(1.0-M_{\rm rem}/M_{\star})$. $M_{\rm rem}/M_{\star}$ is given in Table~\ref{tab:boundunbound}. The diagonal dotted line in each panel represents $dM/dE\propto e^{-k |E|/\Delta\epsilon}$ with $k=4.0$ (\textit{left} panel) and $7.5$ (\textit{right} panel). }
	\label{fig:dmde_PD}
\end{figure*}

\section{Simulations}\label{sec:sim_setup}

We treated stellar masses $M_{\star} = 0.15$, 0.3, 0.4, 0.5, 0.7, 1.0, 3.0, 10.0.
For each, we ran a set of simulations with pericenters $r_{\rm p}$ chosen so as to span the range from total disruptions to weakly partial disruptions.  These pericenters may be described in terms of the order-of-magnitude estimate for the tidal radius $\rtidal$ by writing $r_{\rm p} = \rtidal/\beta$, where $\beta$ is the so-called ``penetration factor''. The largest pericenter studied was chosen so that mass lost from the star was several percent of the star's initial mass.

We distinguish full from partial disruptions by three conditions: 
\begin{enumerate}
\item \label{con1} Lack of any approximately-spherical bound structure. 
\item \label{con2} Monotonic (as a function of time) decrease in the maximum pressure of the stellar debris. 
\item \label{con3} Monotonic decrease in the mass within the computational box. The mass remaining in the box for complete disruption falls with increasing distance from the BH $\propto r^{-\alpha}$ with $\alpha\simeq1.5-2.0$, whereas for partial disruptions the remaining mass eventually becomes constant, which signifies a persistent self-gravitating object. 
\end{enumerate} 	
Events violating any one of these conditions we deem ``partial"; in all cases, if one is violated, all are.

We estimate the physical tidal radius $\physrad$ as the mean of the largest $r_{\rm p}$ yielding a full disruption and the smallest $r_{\rm p}$ producing a partial disruption. 
As shown in \citetalias{Ryu1+2019}, for $M_{\rm BH}=10^{6}$, $\mathcal{R}_{\rm t}/\rtidal \simeq 1$--1.4 for low-mass stars ($0.15\leq M_{\star}\leq0.5$);  falls rapidly between $M_{\star}\simeq0.5$ and $1.0$; and is roughly constant at $\simeq 0.45$ for high-mass stars ($M_{\star}\geq1$). As a result, for stars with $0.15 \leq M_{\star} \leq 3$, all orbits with $r_{\rm p}\gtrsim 27~r_{\rm g}$ lead to at most partial disruption. Here, $r_{\rm g}=G M_{\rm BH}/c^{2}$ refers to the gravitational radius of the BH. 

Figure~\ref{fig:overview} shows the evolution of the density distribution of a $1\Msol$ star when it is partially disrupted as it traverses an orbit with $r_{\rm p} = 0.55\rtidal = 1.16~\mathcal{R}_{\rm t}$.
Note how it begins to stretch shortly before reaching pericenter, but continues to lose mass until it swings out to $\gtrsim 10\rtidal$.

For the partial disruptions discussed in this paper, we followed the progress of the event until the remnant reached distances from the black hole $\gtrsim 20\rtidal$, equivalent to a time past pericenter $\gtrsim 30\times$ the initial star's vibrational time. The precise distance at which we stopped the simulation was determined by the point at which the remnant mass ceased changing.

\section{Results}\label{sec:results}

Partial tidal disruptions produce two distinct products: a remnant and gaseous debris.  The debris resembles that of full disruptions in the sense that roughly half is unbound and half is bound to the black hole.  The bound debris can return to the black hole, generating a bright flare.  On the other hand, there is a remnant, of course, only in a partial disruption.

\subsection{Stellar debris - Distribution of specific energy and fallback rate}

\label{subsub:dmde_partial}

The most observationally-significant property of the debris is its energy distribution $dM/dE$. This quantity determines the fallback rate of bound debris and the ejection speeds of unbound debris.  \citet{Lacy+1982} pointed out that there is a characteristic scale for the energy of tidal disruption debris,
\begin{equation}
    \Delta \epsilon \sim \frac{GM_{\rm BH} R_{\star}}{\rtidal^2},
\end{equation}
and the distribution $dM/dE$ should be roughly symmetric around $E=0$.

We measure $dM/dE$ by continuously adding up the mass and energy of each fluid element leaving the simulation box. For this purpose, we define $E$ as the relativistic specific orbital energy evaluated in the BH frame, minus rest mass energy.  It is a well-defined quantity because, for all but the final $\lesssim 0.5\%$ of mass-loss, very nearly all the gas leaves the simulation box unbound to the remnant (the bound fraction is $\lesssim 10^{-4}$).  Because we employ a simulation box elongated in the direction of debris flow and most of the work done on the gas by the remnant's gravity happens when the gas is relatively close, we capture most of the change in energy due to this effect (see, e.g., \citealt{Guillochon+2013}).  Put another way, the box is long enough that it contains the remnant's Hills radius until roughly the end of the simulation, and by this point the overwhelming majority of mass lost has traveled far outside the Hills radius.

The finite size of the box may, however, lead to a small overestimate of the orbital energy of unbound gas and a similarly small underestimate of the energy of bound gas.  The fractional error is $\sim \langle \cos\theta \rangle (\Delta \epsilon/\Delta E) (R_{\star}/L_x)^{1/2}(r_{\rm t}/\langle r\rangle)^{1/2}$, which is $\simeq 0.05$ for typical parameter values.  Here $\theta$ is the angle between the line connecting a debris fluid element to the remnant and the velocity of the fluid element, $\Delta E$ is the characteristic scale of the energy distribution, $L_x$ is the size of the box in its long dimension, and $\langle r\rangle$ is the mean distance of the star from the black hole when the mass is lost.

In the \textit{left} panel of Figure~\ref{fig:dmde_PD}, we show $dM/dE$ for the stellar debris produced by severe partial disruptions. By ``severe", we mean events in which the remnant mass $M_{\rm rem}/M_{\star}\simeq40-60\%$.  These events have pericenters not much greater than $\mathcal{R}_{\rm t}$ ($r_{\rm p}/\mathcal{R}_{\rm t}\simeq 1.2$).  The \textit{right} panel of Figure~\ref{fig:dmde_PD} shows $dM/dE$ for ``weak" partial disruptions, those in which $M_{\rm rem}/M_{\star}\gtrsim90\%$ and $r_{\rm p}/\mathcal{R}_{\rm t}\simeq 1.5-2.0$.
Because our sample was bimodal in terms of mass-loss (only 3 of our 32 cases had fractional mass-loss between 10\% and 40\%), these two extremes comprise most of the cases we studied.

\begin{figure*}
	\centering
	\includegraphics[width=8.9cm]{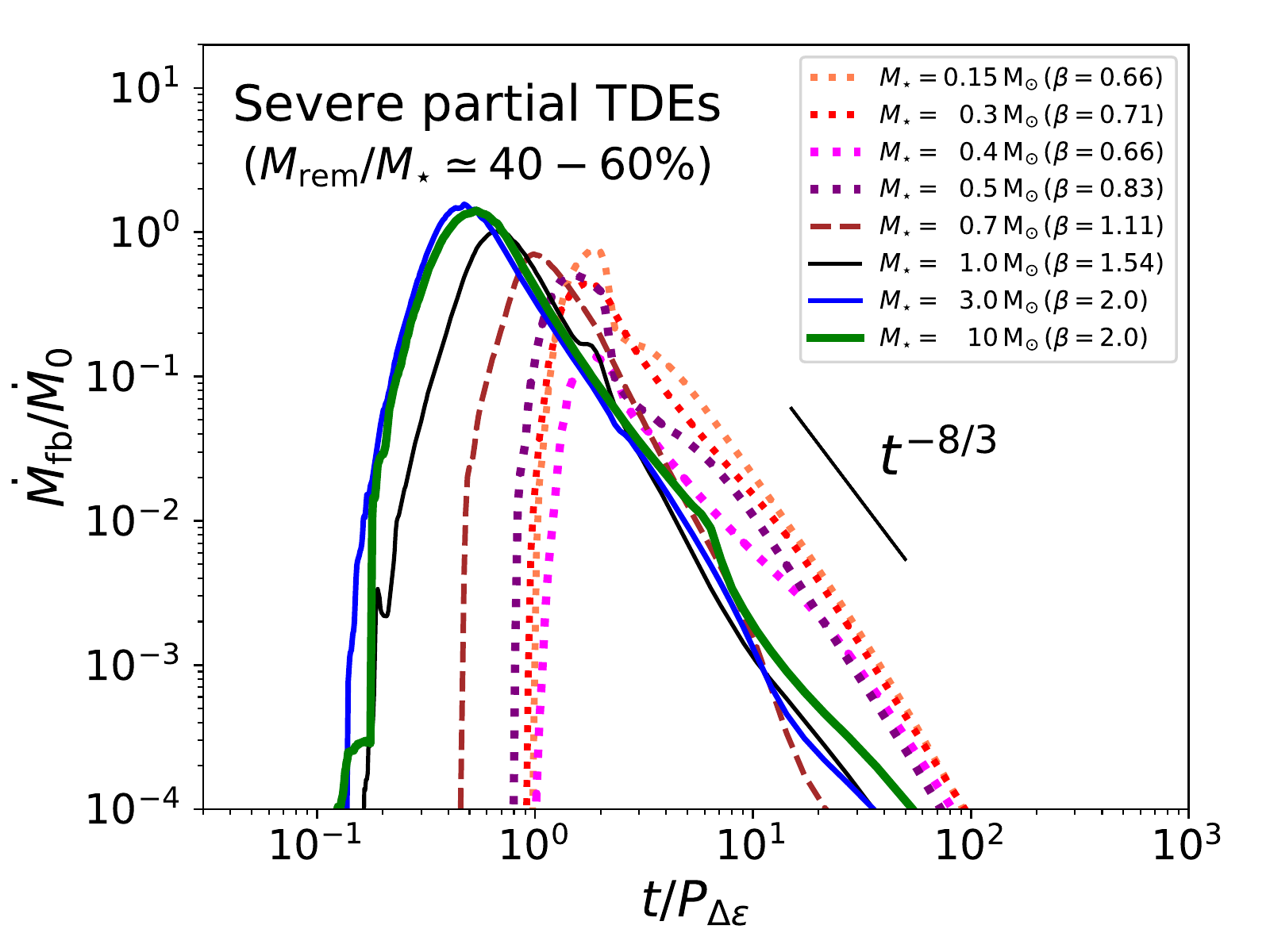}		
	\includegraphics[width=8.9cm]{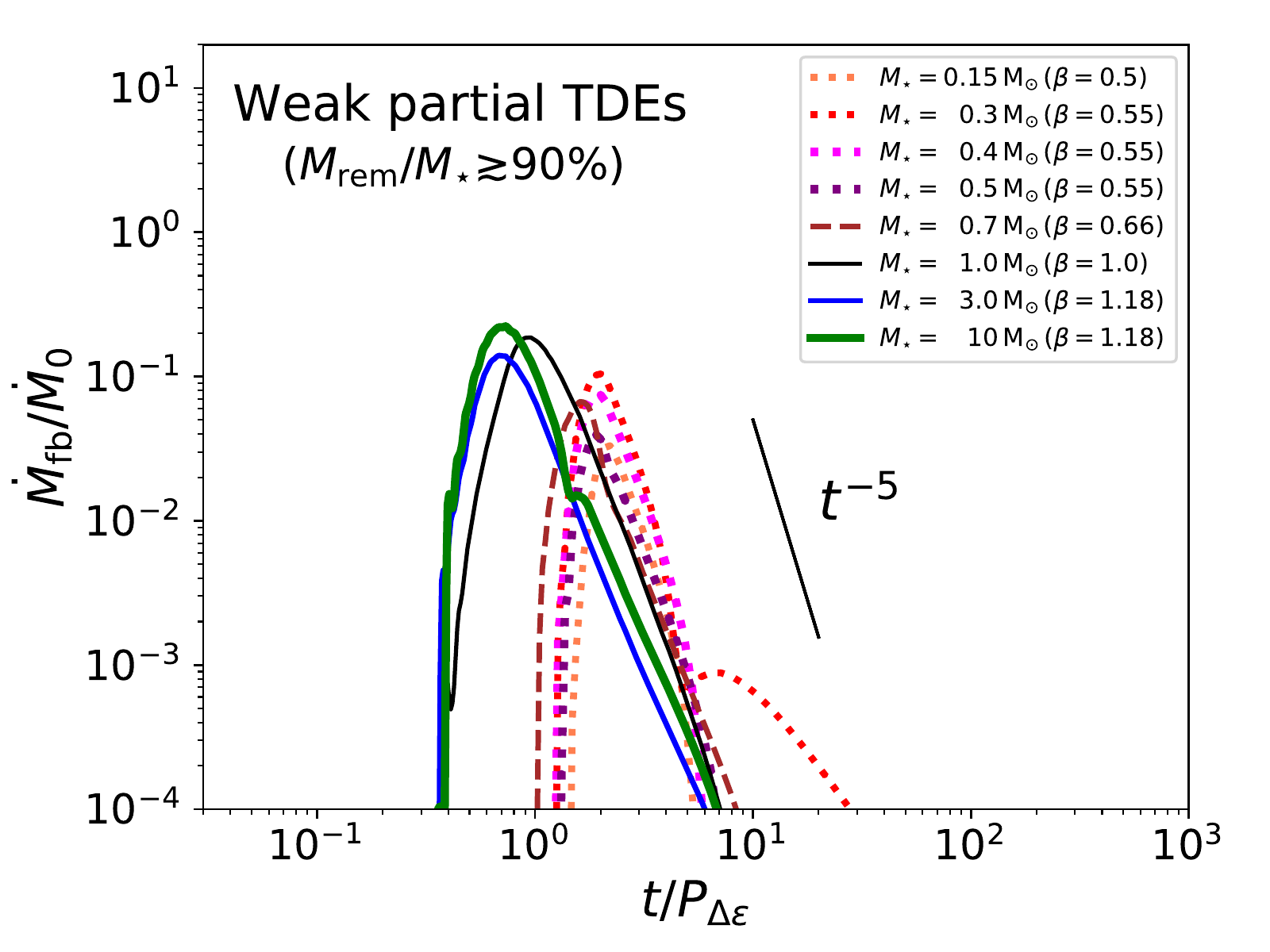}		
	\caption{The fallback rate $\dot{M}_{\rm fb}$ for partial TDEs using the energy distribution in Figure \ref{fig:dmde_PD}. We normalize the time $t$ by the orbital period $P_{\Delta\epsilon}$ and the fallback rate $\dot{M}_{\rm fb}$ by $\dot{M}_{0}=M_{\star}/(3P_{\Delta\epsilon})$. The diagonal solid lines show the power-law $t^{-p}$ with $p=8/3$ (\textit{left} panel) and $p=5$ (\textit{right} panel). The fractional mass of the debris bound to the BH is $\simeq 0.5(1.0-M_{\rm rem}/M_{\star})$, and $M_{\rm rem}/M_{\star}$ is given in Table~\ref{tab:boundunbound}.}
	\label{fig:fallback}
\end{figure*}

 As we showed in \citetalias{Ryu2+2019}, explicit calculations find that the actual distribution $dM/dE$ in complete disruptions is, indeed, very symmetric as \citet{Lacy+1982} predicted, but the magnitude of the energy is correct only at the order of magnitude level. The characteristic spread in energy $\Delta E$, defined as the energy width containing 90\% of the total mass, is $\simeq 0.8 \Delta\epsilon$ for low-mass stars ($0.15\leq M_{\star}\leq0.5$), but jumps to $\simeq 1.5\Delta\epsilon$ for $M_{\star}\approx 1$ and rises to almost 2 for higher-mass stars.  For all masses, $dM/dE$ has local maxima at $E \simeq \pm \Delta E$, but drops smoothly toward $E \approx 0$, where there is a local minimum whose value is only $\simeq 2/3$ that found at the maxima. In low-mass stars, $dM/dE$ plummets for $|E| > \Delta E$; in high-mass stars, it falls exponentially toward larger $|E|$, but on a scale $\simeq \Delta\epsilon/3$, so that there can be a noticeable amount of mass in the wings.

As shown in Figure~\ref{fig:dmde_PD}, some of these characteristics are replicated in partial disruptions, but with the notable contrasts that the local minimum near $E=0$ is much deeper, and $\Delta E$ is a function of $r_{\rm p}/\mathcal{R}_{\rm t}$ as well as of $M_{\star}$.  Not too surprisingly, in severe partial disruptions $\Delta E$ is consistently close to its value in full disruptions.   However, it drops by a factor $\simeq 2$ going from severe disruptions to weak ones.  Severe disruptions also resemble full disruptions in that $dM/dE$ for high-mass stars, but not low-mass stars, has exponential wings.  These differ, however, in that they are somewhat steeper: $dM/dE \propto e^{-4|E|/\Delta\epsilon}$ rather than $\propto e^{-3 |E|/\Delta \epsilon}$.  In weaker partial disruptions, the exponential wings decline more rapidly, on scales a factor $\sim 2$ shorter than in the severe cases.

The greatest contrast between partial disruptions and full disruptions is in the depth of the central minimum.  The factor $\simeq 2/3$ between $dM/dE(E=0)$ and $dM/dE(E=\Delta E)$ for full disruptions becomes a factor $\sim 10^{-2}$ for partial disruptions.  The very deep central minimum results in  nearly all the debris mass being concentrated near $E \simeq \pm \Delta E$.

In Figure~\ref{fig:fallback}, we show the fallback rate for the two partial disruption cases, calculated using the energy distributions shown in Figure~\ref{fig:dmde_PD} and the expression for the fallback rate \citep{Rees1988,Phinney1989},
\begin{align}
\dot{M}_{\rm fb}=\left(\frac{M_{\star}}{3P_{\Delta\epsilon}}\right)\left( \frac{dM/M_{\star}}{d\epsilon/2\Delta\epsilon}\right)\left(\frac{t}{P_{\Delta\epsilon}}\right)^{-5/3},
\label{eq:mdot2}
\end{align} 
where $P_{\Delta\epsilon}= (\uppi/\sqrt{2})G M_{\rm BH}\Delta\epsilon^{-3/2}$ is the orbital period for orbital energy $-\Delta\epsilon$. The most noticeable feature is greater deviations from the $t^{-5/3}$ power-law for weaker tidal encounters. This effect is directly due to the progressively smaller amount of mass with $E \simeq 0$ as the events weaken.  Even for the severe events, however, the decline is noticeably steeper than $t^{-5/3}$.  As shown in the left panel of Figure~\ref{fig:fallback}, the slope is $\simeq -2.7$ for the high-mass stars, and somewhat shallower for low-mass stars (between $\simeq -2$ and $\simeq -2.7$).   For weak events, the fallback rate declines fastest for the low-mass stars ($\propto t^{-6}$) and a bit more gently for the high-mass stars ($\propto t^{-5}$).   These power-laws are best-determined for times when ${\dot M}_{\rm fb}/{\dot M}_0 \gtrsim 10^{-3}$; the total mass returning at later times is so small that it could radiate very little energy.
As is true of total disruptions, the peak in the fallback rate for low-mass stars is both sharper than for high-mass stars and delayed by factor $\simeq 3$; these contrasts directly reflect the narrower energy width in the debris from low-mass stars (Figure~\ref{fig:dmde_PD}).

These results bear a qualitative resemblance to those of \citet{Guillochon+2013}, but also disagree in some aspects. Direct comparison is possible only for their $M_{\star}=1$ polytrope with $\gamma=4/3$. In both their calculations and ours, the slope of the decline is greater for weaker events.  However, in their case the contrast is substantial only for the first $\sim 3-5$ $P_{\Delta\epsilon}$, after which the logarithmic slope for the weakest encounters, whose most negative value is $-3.7$, becomes as shallow as $\simeq -2.3$ (see their Figure~7).  By contrast, our $M_{\star}=1$ results show a fairly constant power-law slope $\simeq -2.7$  for severe disruptions up to the point at which ${\dot M}_{\rm fb}/{\dot M}_0$ falls below $10^{-3}$ (at $\simeq 10 P_{\Delta\epsilon}$) and a similarly constant power-law slope $\simeq -5$ up to the same fallback rate cut-off for a weaker one.  Some of these contrasts may be due to our coarser sampling in $\beta$; however, especially for weak partial disruptions, a more important source of contrast may be the differing density profiles in the outer portions of $M_{\star}=1$ stars predicted by a realistic density profile and a $\gamma=4/3$ polytrope (see Figure~2 in \citetalias{Ryu2+2019}).

Our results also conflict with the claim of \citet{CoughlinNixon2019} that the post-peak logarithmic slope $p$ for partial disruptions gradually steepens to an asymptote of $\simeq 9/4$ independent of $M_{\rm rem}$, owing to a continuous gravitational influence of the remnant on the debris marginally bound to the BH. Several methodological contrasts may account for this disagreement. Whereas we use a full $3-$dimensional hydrodynamic simulation to describe the complex geometry of the tidal streams and remnant, \citet{CoughlinNixon2019} use a $1-$dimensional analytic model in which both the debris streams and the remnant move exclusively in the radial direction with respect to the black hole.  This assumption has the consequences that the gravitational force exerted by the remnant on a gas parcel is purely radial, and its magnitude is determined by the difference between their distances from the black hole.  It also implies that the work done by the remnant on the fluid elements does not reflect any obliquity between the direction of motion of the fluid and the direction between it and the remnant.  Finally, whereas we compute the self-gravity of both the mass in the stellar remnant and the debris contained within a large box around the remnant ($17~R_{\star} \times 9~R_{\star} \times 14~R_{\star}$), \citet{CoughlinNixon2019} ignore the self-gravity of the debris.  Our approach accurately calculates the work done on the fluid by the remnant while it remains within the simulation box; because the total amount of work is dominated by the portion done while the fluid element is nearest the remnant, our box is large enough to account for the majority of this effect.

\citet{Golightly+2019} presented one example of a partial TDE  taking place in a star directly comparable to one of ours: a $3\Msol$ star whose structure was computed with \mesa and was halfway through its main-sequence lifetime.  Using the SPH code {\small PHANTOM}, they found a fallback rate exhibiting a late-time slope $\simeq -9/4$.    The pericenter for this encounter, $r_{\rm p} = 0.33 \rtidal$, was, however, smaller than $\physrad$ as determined by our simulations ($\simeq 0.4-0.45\rtidal$).  It is possible that they found only a partial disruption, but perhaps a rather strong one, because they employed Newtonian rather than relativistic gravity, even though this pericenter is only $27~r_{\rm g}$.

\citet{Goicovic+2019} also studied the shape of the debris energy distribution  and the consequent fallback rate for a $M_{\star}=1$ star whose initial mass profile was taken from \mesa~data.  Comparing their $\beta=1.6$ and $\beta=1.1$ cases with ours having $\beta = 1.54$ and $\beta = 1.0$, we find  (comparing to their Figure~5a) good consistency:  from the time of peak fallback rate to a time $10\times$ greater, we both find a mean slope $\simeq -2.5$ in the former case and $\simeq -3$ in the latter. Similarly to ours, the $dM/dE$ distribution in their Figure~4 shows the appearance of wings near the outer boundaries, and these wings become steeper for weaker encounters. Given the consistency in $dM/dE$, it is not surprising to find similar fallback rates as well.

\begin{figure}
	\centering
	\includegraphics[width=9.1cm]{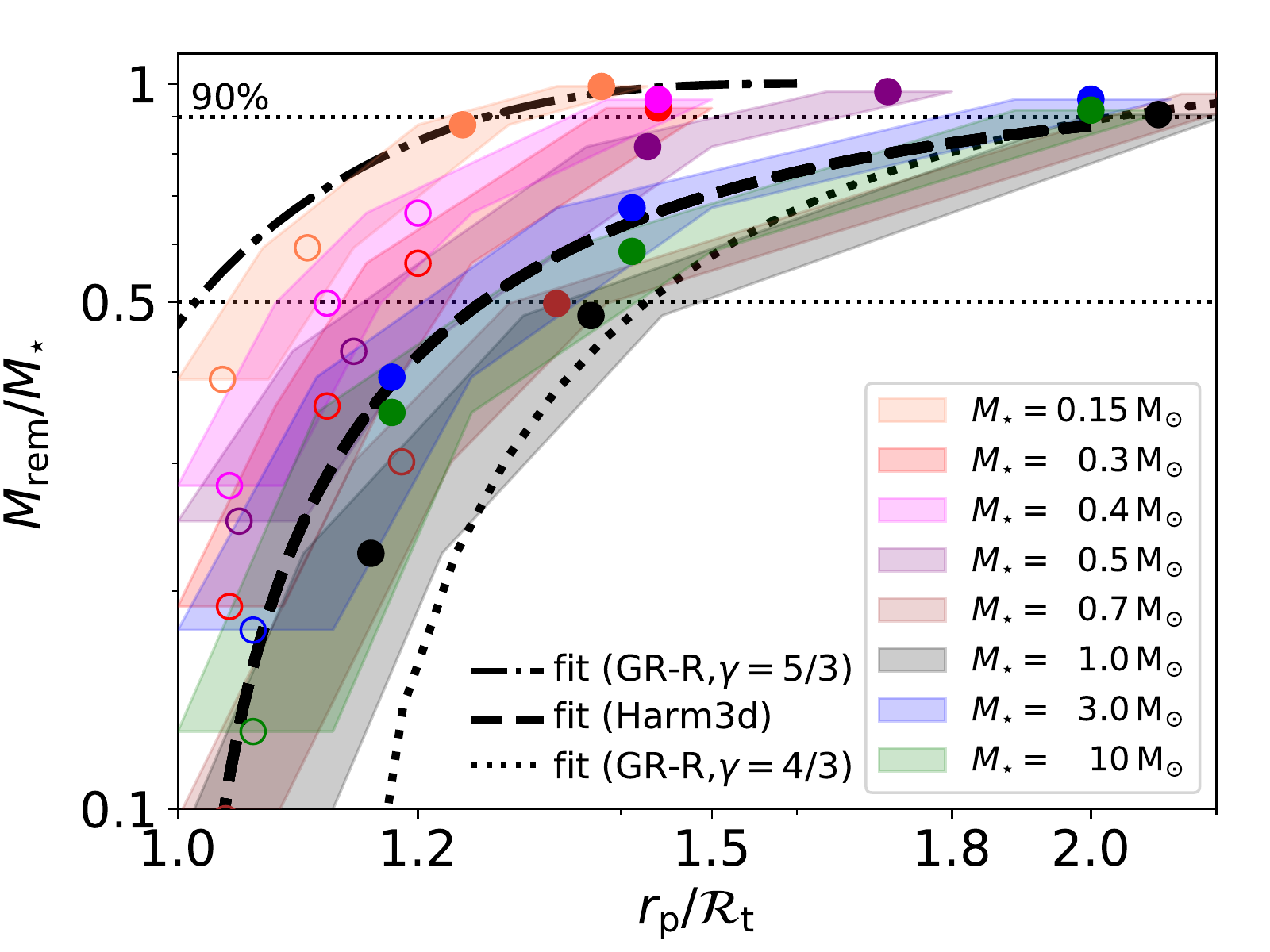}			
	\caption{The fractional remnant mass $M_{\rm rem}/M_{\star}$ as a function of pericenter distance normalized to physical tidal radius, i.e., $r_{\rm p}/\mathcal{R}_{\rm t}$. The shaded regions around the solid lines demarcate the ranges determined by the uncertainties of $\mathcal{R}_{\rm t}$, filled with the same colors as the solid lines. The uncertainty in $\mathcal{R}_{\rm t}$ is due to our discrete sampling of $r_{\rm p}$ ($0.05-0.1$ in $r_{\rm p}/r_{\rm t}$). The dotted horizontal lines show the 50\% and 90\% remnant mass-fraction levels. 
	The fitting formula given in Equation \ref{eq:bestfit_remmass1} is plotted using a thicker black dashed line. The fitting formulae for $1\Msol$ polytropic stars with $\gamma=5/3$ and $\gamma=4/3$ by \citet{Guillochon+2013} (GR-R) are depicted using thinner dot-dashed and dotted curves, respsectively. The circle markers indicate whether each remnant has a positive (unfilled) or negative (filled) orbital energy with the BH potential.}
	\label{fig:remnant_mass}
\end{figure}

\subsection{Surviving remnants}\label{subsec:remnant}

\subsubsection{Mass}
\label{subsub:mass_remnant}

Figure~\ref{fig:remnant_mass} shows the fractional remnant mass $M_{\rm rem}/M_{\star}$ as a function of $r_{\rm p}$.
When low-mass stars have $r_{\rm p}\gtrsim 1.5\mathcal~\mathcal{R}_{\rm t}$, even though they are tidally deformed near the pericenter, they recover their (quasi-) spherical structures without a significant loss of mass ($\lesssim10\%$). For high-mass stars, such weak mass-loss occurs for $r_{\rm p}\gtrsim 1.8~\physrad$.

We find that a simple functional form,
\begin{align}
\label{eq:bestfit_remmass1}
\frac{M_{\rm rem}}{M_{\star}}  &= 1.0 - \left(\frac{r_{\rm p}}{\mathcal{R}_{\rm t}}\right)^{-3.0},
\end{align}
captures the key features of the pericenter-dependence of $M_{\rm rem}/M_{\star}$.  In fact, by coincidence, it reproduces the curve for $M_{\star}=3$ almost exactly.
\citet{Guillochon+2013} also provides fitting formulae for the remnant mass of polytropic stars with $\gamma=4/3$ and $5/3$ ($1.0-C_{\gamma}$ in their Appendix), as a function of $\rtidal/r_{\rm p}$. Their formulae for these two values of $\gamma$ run along the envelope of the remnant mass curves shown in Figure~\ref{fig:remnant_mass}: the curve for $\gamma=5/3$ lies slightly above that for $M_{\star}=0.15$, while the curve for $\gamma=4/3$ is close to that for $M_{\star}=1$.  In other words, compared to our calculations for fully convective low-mass stars, their remnants retain greater mass, while compared to our calculations for $M_{\star}=1$ stars, there is reasonable agreement.

Although it is remarkable that such a simple expression can well characterize the remnant mass, Equation~\ref{eq:bestfit_remmass1} does not attempt to describe $M_{\rm BH}$-dependence of $M_{\rm rem}/M_{\star}$. \citetalias{Ryu1+2019} shows that when $r_{\rm p}/r_{\rm t}$ is rewritten in terms of the specific orbital angular momentum in units of $r_g c$ (Equation~11 in \citetalias{Ryu1+2019}), it becomes valid {\it independent} of $M_{\rm BH}$.

In \citetalias{Ryu2+2019}, we introduced a semi-analytic model which predicts {the physical tidal radius $\physrad$ and the maximum radius at which significant mass can be removed in a partial disruption $\widehat R_{\rm t}$ on the basis of the star's central density and mean density, respectively.} 
By combining the empirical Equation~\ref{eq:bestfit_remmass1} and this semi-analytic model, we can obtain a direct relationship between three dimensionless spatial scales, i.e., $\mathcal{R}_{\rm t}/r_{\rm t}(=\Psi)$, $r_{\rm p}/r_{\rm t}(=\beta^{-1}\geq\Psi)$ and $R/R_{\star}$, the fractional radius within the star containing $M_{\rm rem}$. Inserting $M_{\rm rem}/M_{\star}$ from Equation~\ref{eq:bestfit_remmass1} into the equation defining this model's basic assumption (that lost mass is taken from outside the point at which the tidal gravity matches an empirically-determined multiple of the star's self-gravity), we find:
\begin{align}
\frac{R}{R_{\star}} \simeq 0.47 \left(\left[\frac{r_{\rm p}}{r_{\rm t}}\right]^{3}-\left[\frac{\mathcal{R}_{\rm t}}{r_{\rm t}}\right]^{3}\right)^{1/3}.
\label{eq:connection}
\end{align} 
This relation behaves correctly in simple limits: at $\beta^{-1} = \Psi$, $R=0$, and at $\beta^{-1}=\widehat{R}_{\rm t}/r_{\rm t}$ ($\widehat{R}_{\rm t}$ the largest pericenter distance for tidal mass-loss, see Equation 17 in \citealt{Ryu2+2019}), $R/R_{\star}\simeq1$ with no more than 5\% errors.  Thus, with a model for the star's initial mass profile and knowledge of $\physrad$, the remnant mass can be predicted easily for any pericenter larger than the physical tidal radius.

	\begin{table*}
		\renewcommand{\thetable}{\arabic{table}}
		\centering
		\caption{The properties of partial disruption remnants.  In the left-hand columns, we list the original mass of our model stars $M_{\star}~[\rm{M}_{\odot}]$, $r_{\rm t}/r_{\rm p}(\equiv\beta)$, $r_{\rm p}/\mathcal{R}_{\rm t}$, the remnant mass $M_{\rm rem}~[\rm{M}_{\odot}]$, the mass fraction $M_{\rm rem}/M_{\star}$, the sign of the mass-weighted specific energy ${\bar E}$ (B:${\bar E}<0$ and U:${\bar E}>0$) and the magnitude of the average specific energy in units of $\Delta \epsilon$.
		The right-hand four columns give orbital parameters for the remnants: for unbound stars, only the ejection velocity $v_{\rm ejec}$; for bound stars, the eccentricity $\bar{e}$, semimajor axis $a$ and orbital period $P$. The orbital parameters and the remnant mass are measured when those quantities have settled into asymptotic values (at $r \simeq20-30\rtidal$). Note that we do not show $v_{\rm ejec}$ for the $M_{\star}= 3$ and $10$ cases' most severe disruptions.  This is because even at $r \simeq20-30\rtidal$, they had not settled into an approximate steady state; in addition, their mean specific energy was so different from that of the initial star's that the remnant was offset far enough from the center of the simulation box that some of its mass was no longer inside the box. We exclude these two cases from the analysis of the unbound population in the text.}
		\label{tab:boundunbound}
		\begin{tabular}{c c c c  c c c c | c c c c  }
			\tablewidth{0pt}
			\hline
			\hline
			$M_{\star}$&  $ r_{\rm t}/r_{\rm p}(\equiv\beta)$ & 	$r_{\rm p}/\mathcal{R}_{\rm t}$& $M_{\rm rem}$ &  		$M_{\rm rem}/M_{\star}$ & 	B/U  & 	$\log_{10}(|\bar{E}|/\Delta\epsilon)$ &$v_{\rm ejec}[{\rm km}\s^{-1}$] & 	$\log_{10}(1-\bar{e})$ &  $a~$[pc]& 	$P~[10^{3}\yr]$\\
			\hline
			\multirow{4}{*}{$0.15$}  
& 0.50 &  1.38  &  0.14  & 0.99  &  B & -2.9  & -  &  -4.6  &  0.058  &  1.3 \\
& 0.56 &  1.24  &  0.13  & 0.87  &  B & -3.2  & -  &  -4.9  &  0.11  &  3.4 \\
& 0.63  &  1.10 & 0.08  & 0.59  & U & -3.3  & 174  & - & -  & -  \\
  & 0.67 &  1.03  &  0.05  & 0.39  & U & -2.9  & 295  & - & - &  -\\
\hline
			
			\multirow{4}{*}{$0.3$}  
			& 0.56 &  1.44 &  0.28  & 0.92   &  B & -3.4    & -&  -5.0  &  0.18  &  7.3 \\
			 & 0.67 &  1.20 &  0.17  & 0.56 &  U & -3.8  & 94 & -& - & -  \\
			 & 0.71 &  1.12  &  0.11  & 0.36&   U & -3.2 &180& - &  - & -  \\
			& 0.77 &  1.04 &  0.06  & 0.18 & U & -3.7 &110&- &  - & -  \\
			\hline
						\multirow{4}{*}{$0.4$}  
 			 & 0.56 &  1.44  &  0.38  & 0.95  &  B & -3.2  & -  &  -4.7  &  0.11  &  3.6\\
 			& 0.67 &  1.20  &  0.26  & 0.66  & U & -3.7  & 107  & - & - &  -  \\
			& 0.71 &  1.12  &  0.19  & 0.49  & U & -3.2  & 193  & - & - &  - \\
			 & 0.77 &  1.04  &  0.11  & 0.27  & U & -2.7  & 334  & - & - &  -  \\
			\hline
			
			\multirow{4}{*}{$0.5$} 
		      & 0.56 &  1.71 &  0.49  & 0.97  &B & -2.9    & -    &  -4.5&  0.071  &  1.8 \\
			 & 0.67 &  1.43   &  0.41  & 0.81 &B & -3.8      & - &  -5.4 &  0.50  &  33 \\
			 & 0.83 &  1.14  &  0.22  & 0.43&  U & -3.8  & 93&-   &  - & -\\
			 & 0.91  &  1.05  &  0.13  & 0.24&  U & -3.4  &  150  &-& - & -  \\
			\hline
			\multirow{4}{*}{$0.7$}  
             & 0.67 &  2.22  &  0.67  & 0.96  &  B & -3.1  & -  &  -4.6  &  0.11  &  3.5 \\
             & 1.11 &  1.33  &  0.34  & 0.49  &  B & -4.2  & -  &  -6.0  &  1.7  &  200\\
            & 1.25  &  1.19  &  0.21  & 0.30  & U & -2.6  & 322  & - & - &  - \\
             & 1.43  &  1.04  &  0.06  & 0.09  & U & -2.7  & 299  & - & - &  -\\
            \hline	
			\multirow{4}{*}{$1.0$} 
			& 1.00 &  2.11 &  0.91  & 0.91 &  B & -2.9  & -  &  -4.6  &  0.087  &  2.4 \\
			 & 1.54 &  1.37   &  0.48  & 0.48 &  B & -2.6   & - &  -4.5  &  0.047 &  0.97  \\
			 & 1.82 &  1.16 &  0.23  & 0.22 &  B & -2.4     & -    &  -4.4 &  0.028  &  0.44  \\
		 & 2.00 &  1.05   &  0.08  & 0.08&B & -2.9    & - &  -4.9  &  0.088  &  2.5 \\
			\hline
			\multirow{4}{*}{$3.0$} 
			 & 1.18 &  2.00  &  2.85  & 0.95& B & -2.8     & -  &  -4.4 &  0.083  &  2.2\\
			 & 1.67 &  1.41  &  2.02  & 0.67  &  B & -3.6    & -  &  -5.4   &  0.56  &  40\\
		 & 2.00  &  1.18 &  1.18  & 0.39 &   B & -2.9        & - &  -4.7 &  0.10  &  3.0  \\
		& 2.22 &  1.06 &  0.53  & 0.17  & U & -2.3  &  * &  * & * & * \\
			\hline
			\multirow{4}{*}{$10$} 
		 & 1.18 &  2.00   &  9.19  & 0.91 &  B & -3.0     & - &  -4.4&  0.13  &  4.5  \\
		 & 1.67 &  1.41   &  5.87  & 0.58 &   B & -2.9  & -  &   -4.4  &  0.097  &  2.8 \\
			& 2.00 &  1.18   &  3.52  & 0.35&   B & -2.9      & -&  -4.6   &  0.11  &  3.6 \\
			 & 2.22  &  1.06  &  1.28  & 0.12 &  U & -1.7   & * & * & * &  * \\
			\hline
		\end{tabular}
	\end{table*}

\begin{figure*}
	\centering	
	\includegraphics[width=9.1cm]{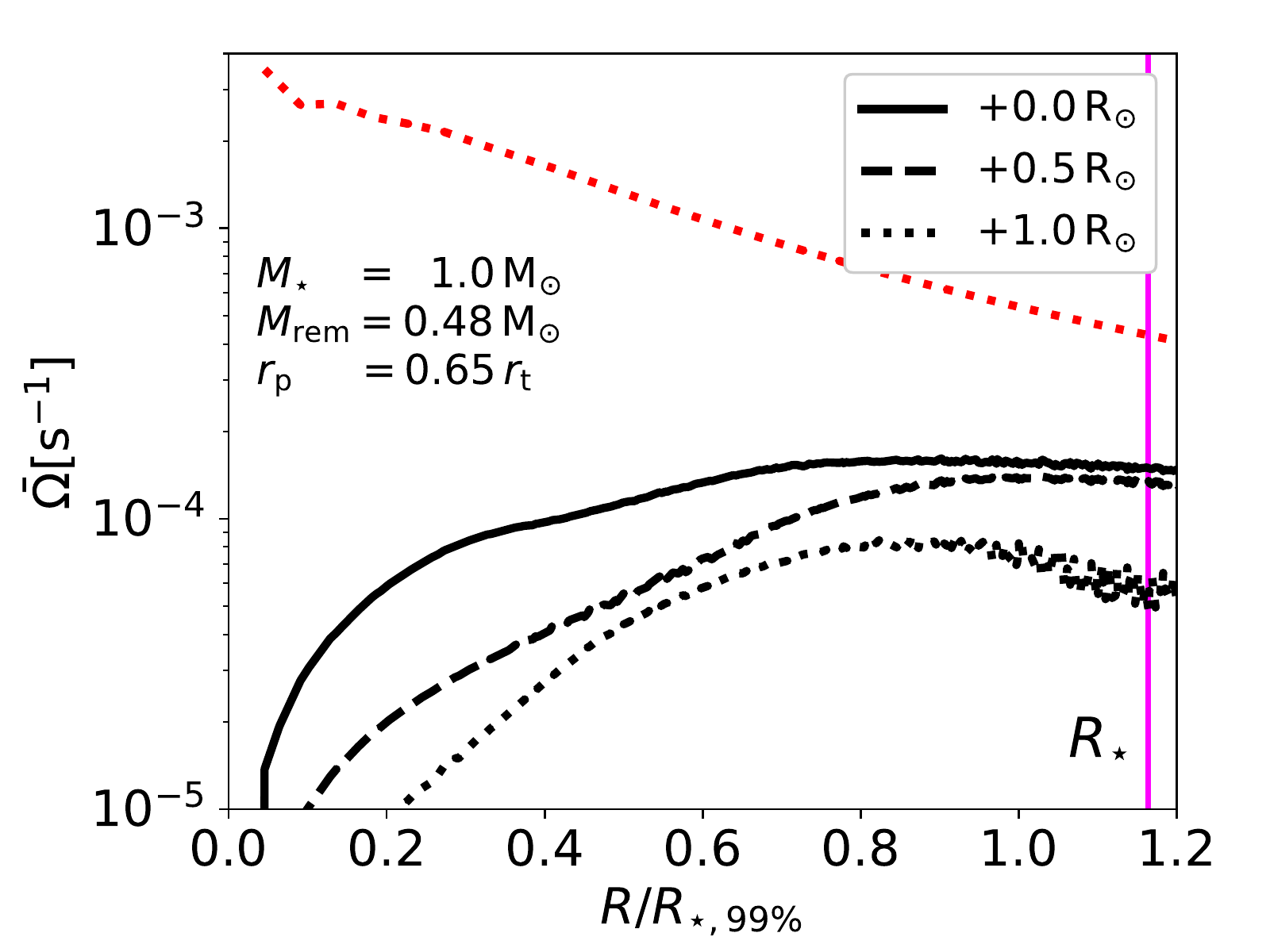}				\hspace{-0.2in}	
	\includegraphics[width=9.1cm]{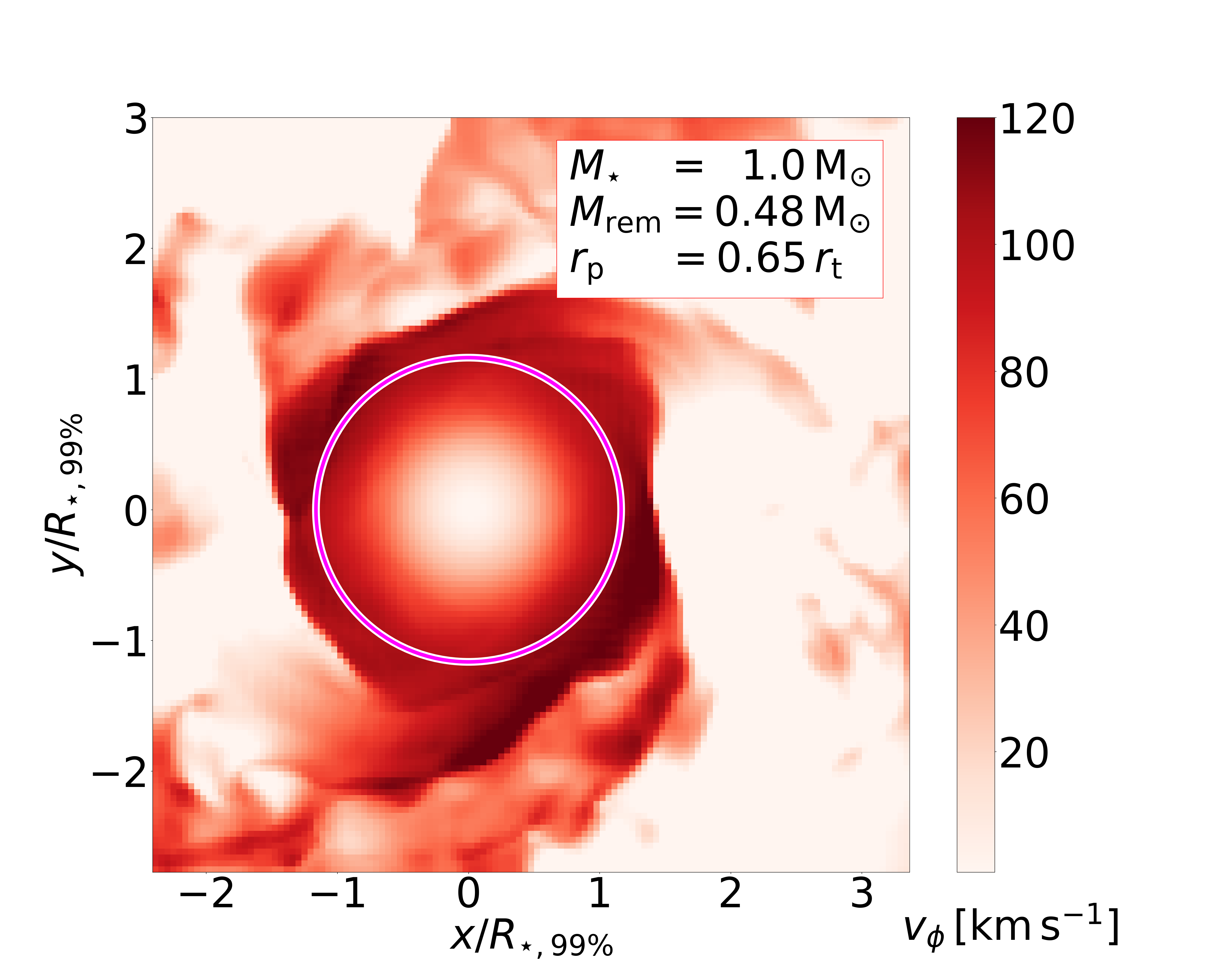}\hspace{6.9cm}
	\caption{Rotational properties of the remnant from an event in which a star with $M_{\star}=1$ passes through a pericenter $r_{\rm p}=0.65\rtidal$. The remnant mass $M_{\rm rem}\simeq0.48$. (\textit{left} panel) The mean angular frequency $\bar{\Omega}(R)$ at cylindrical radius $R$ on three horizontal planes; their heights above the orbital plane are: $z=0\Rsol$ (solid), $0.4\Rsol$ (dashed) and $0.8\Rsol$ (dotted). The radius $R$ on the $x-$axis is normalized by the radius  $R_{\star,99\%}$ for 99\% of the remnant mass. The red dotted line shows the equatorial break-up angular frequency. The vertical magenta solid line at $R/R_{\star,99\%}\simeq 1.16$ is placed at the radius $R_{\star}$ of our $1\Msol$ MS star.
	(\textit{right} panel) Azimuthal velocity $v_\phi$. $x$ and $y$ are normalized by $R_{\star,99\%}$. The solid magenta circle delineates $R_{\star}$. }
	\label{fig:rotation}
\end{figure*}
	
\subsubsection{Specific energy - bound or unbound}
\label{subsub:bound_unboud}

In this section, we focus on the specific energies of surviving remnants to see whether or not they are bound to the BH, and to determine their orbital motion in either case (see Table~\ref{tab:boundunbound} for the results).  We consider the question of whether they are bound to the galaxy's bulge separately.
As a prologue to this topic, it is useful to lay out the hierarchy of orbital energy scales in this problem. The most useful unit for this hierarchy is the specific kinetic energy of stars in the region of the galaxy from which the disrupted stars are drawn, i.e.,  $(1/2)\sigma^{2}$, where $\sigma$ is the $3-$dimensional bulge velocity dispersion.   In terms of this unit, the initial orbital energy of stars in our simulations counting only the black hole's contribution to the gravitational potential is very small, $\sim-10^{-3}(\sigma^2/2)$, which, in relativistic terms, is a specific energy $\sim -10^{-10}c^2$ for $\sigma \sim 100 - 300$~km~s$^{-1}$.

In this sense, one might think of our stars as having, prior to the disruption, energy very close to the middle of the bulge stars' energy distribution.  On the other hand, the magnitude of the typical remnant's specific energy is relatively large, $\sim 1-10$.   Because the typical remnant energy changes by an amount greater than the actual energy with which stars begin the event, we can approximate the remnant's final energy as its actual energy with respect to the BH potential. Moreover, because it is also several times larger than the potential associated with the stars of the inner galaxy, it is appropriate to label remnants with positive final energy as ``unbound" with respect to the innermost portion of the galaxy. However, we must also emphasize that ``large" is a relative term.  Although the remnants' energies are comparable to or larger than the kinetic energy of bulge stars, they are tiny compared to the magnitude of the debris energy, whether bound or unbound---they are $\sim 10^{-3}$ on that scale.

It is a good approximation to suppose that the BH potential dominates the entire region through which bound remnants  travel because all but one of their apocenters ($\simeq 0.05-1\pc$, Table \ref{tab:boundunbound}) are smaller than the BH's radius of influence  ($ \sim 1-10$~pc; see Section~\ref{discuss:unbound} for further discussion of this point). 
 The corresponding periods are between $\simeq 400$ and $\simeq 40,000\yr$.   Their eccentricities are exceedingly close to 1, mostly with $|1-e| \sim 10^{-5}$.
There is also one case ($M_{\star}=0.7$, $r_{\rm p}/\rtidal = 0.9$) that is intermediate between bound and unbound in the sense that it is bound, but only weakly, having $a\simeq1.7\pc$ and $P\simeq0.2\Myr$.  The comparative rarity of remnants whose net energy is very close to zero is likely due to the small associated phase space.
With specific energies similar in magnitude to those of the bound remnants, but opposite sign, the unbound remnants have ejection speeds $v_{\rm ejec}\simeq100-330 \km\s^{-1}$.

Figure \ref{fig:remnant_mass} distinguishes bound from unbound remnants by using filled circles for the former and unfilled circles for the latter.
For low-mass stars, the unbound remnants are associated with the most severe partial disruptions, whereas relatively weak encounters yield bound remnants. However, for high-mass stars, even some severe partial disruptions yield bound remnants.  Because the  specific angular momentum of a remnant (either bound or unbound) is essentially identical to the specific angular momentum of the original star, its pericenter 
(when bound) is very nearly unchanged by the tidal encounter.

A similar studies were reported by \citet{Manukian+2013}.  Using Newtonian hydrodynamics simulations of tidal disruption of polytropic stars  with $\gamma=4/3$,  they determined the orbital energies of remnants at a time $\simeq 100 \sqrt{R_{\star}^{3}/GM_{\star}}$) after pericenter passage.  Contrary to what we found, all their surviving remnants were, in our language, unbound, and their ejection speeds were considerably greater than ours.   For example, in the case of stars with $M_{\star}=1$ (for which a $\gamma=4/3$ polytrope is a reasonable approximation), the ejection speed for their remnants ranged from $\simeq 100$~km/s (for $r_{\rm p}/r_{\rm t} = 1$) to $\simeq 600$~km/s (for $r_{\rm p}/r_{\rm t} = 0.55$).  By contrast, the remnants of our $M_{\star}=1$ simulations with $0.5 \leq r_{\rm p}/r_{\rm t} \leq 1$ were all bound, and the greatest ejection speed we found for any other case was $\simeq 330$~km/s.  It is unclear how to account for these differing results; the difference between relativistic and Newtonian tidal forces might play a part.

\begin{figure}
	\centering
	\includegraphics[width=9.1cm]{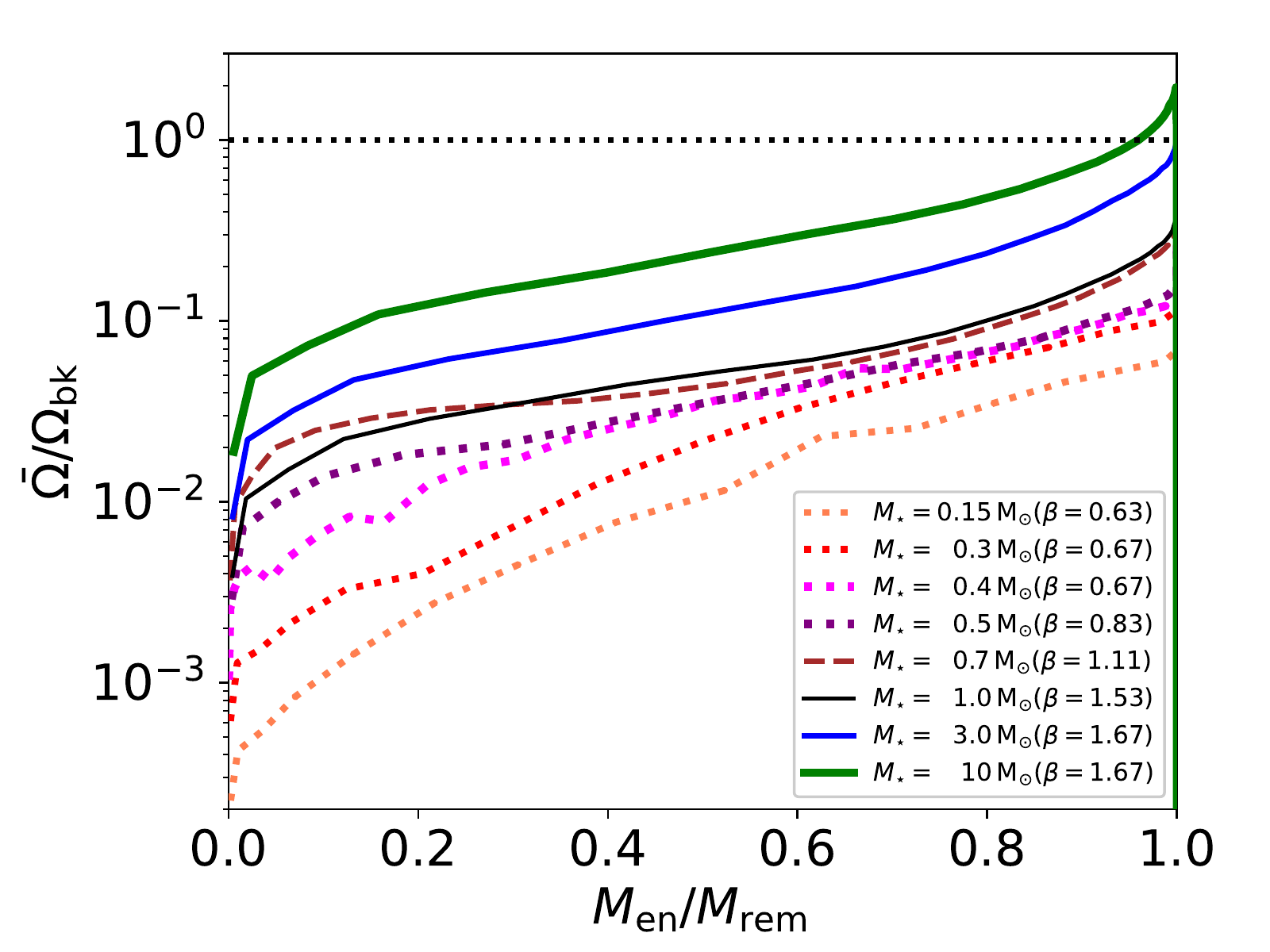}
	\caption{ The ratio $\bar{\Omega}/\Omega_{\rm bk}$ as a function of $M_{\rm en}/M_{\rm rem}$, the ratio of the enclosed mass to the remnant mass. In all cases displayed, $M_{\rm rem}/M_{\star} \simeq 0.5$, or equivalently, $r_{\rm p}/\mathcal{R}\simeq1.1-1.3$.}
	\label{fig:rotation_all}
\end{figure}

\subsubsection{Spin }
\label{res:internal}

All surviving remnants are spun-up in the prograde direction as they are tidally torqued near the pericenter \citep{Rees1988,Goicovic+2019}. As a result, they are approximately oblate spheroids in shape, with the minor axis perpendicular to the orbital plane. In all cases, the angular frequency increases outward. As an example, we present in the \textit{left} panel of Figure~\ref{fig:rotation} the angular frequency $\bar{\Omega}(R)$, an azimuthal average over cells at the same cylindrical radius from an axis through the remnant's center of mass perpendicular to the orbital plane, at three different heights.  The star in this simulation began with mass $M_{\star}=1$, passed through a pericenter $r_{\rm p}=0.65\rtidal$, and emerges from the event with $M_{\rm rem}\simeq0.48$. The angular frequency $\bar{\Omega}$ at each height increases outwards until it reaches a maximum at $R\simeq 0.8-1.0$. The maximum frequency at the equator is around $25-30\%$ of the equatorial break-up angular frequency $\Omega_{\rm bk}$, defined as $\Omega_{\rm bk}(R)=\sqrt{GM_{\rm en}(R)/R^{3}}$. Here $M_{\rm en}(R)$ is the enclosed mass inside cylindrical radius $R$ on the equatorial plane. The rotational velocity $v_{\phi}$ (\textit{right} panel) therefore rises steeply at small radius and then $\propto R$ for $R \gtrsim 0.6\Rsol$.  Its maximum is $\simeq 100-120\km\s^{-1}$.

We find a general trend that, for fixed fractional mass loss, the more massive the initial star, the closer its remnant comes to break-up rotation. This trend is illustrated in Figure~\ref{fig:rotation_all}, in which we present data for partially disrupted stars with $M_{\rm rem}/M_{\star} \simeq 0.5$, corresponding to $r_{\rm p}/\mathcal{R}_{\rm t}\simeq1.1-1.3$. That the high-mass stars reach higher fractions of the break-up rotation rate than the lower-mass stars can be explained simply. To zeroth order, when a star passes through pericenter, tidal forces torque it so that its outer layers rotate at roughly the local orbital frequency. But the local orbital frequency is, by definition, about the same as the vibrational frequency when the distance from the black hole is similar to $r_{\rm t}$. By the same token, the break-up rotational frequency is similar to the vibrational frequency. Consequently, $\Omega/\Omega_{\rm bk} \simeq \Omega(r_{\rm p})/\Omega_{\rm bk} \propto \beta^{3/2}$.  It is also worth noting that if the star spins at near break-up rates {\it before} the encounter, tidal dynamics can be quite different \citep{SacchiLodato2019}.

\begin{figure}
	\centering
	\includegraphics[width=8.9cm]{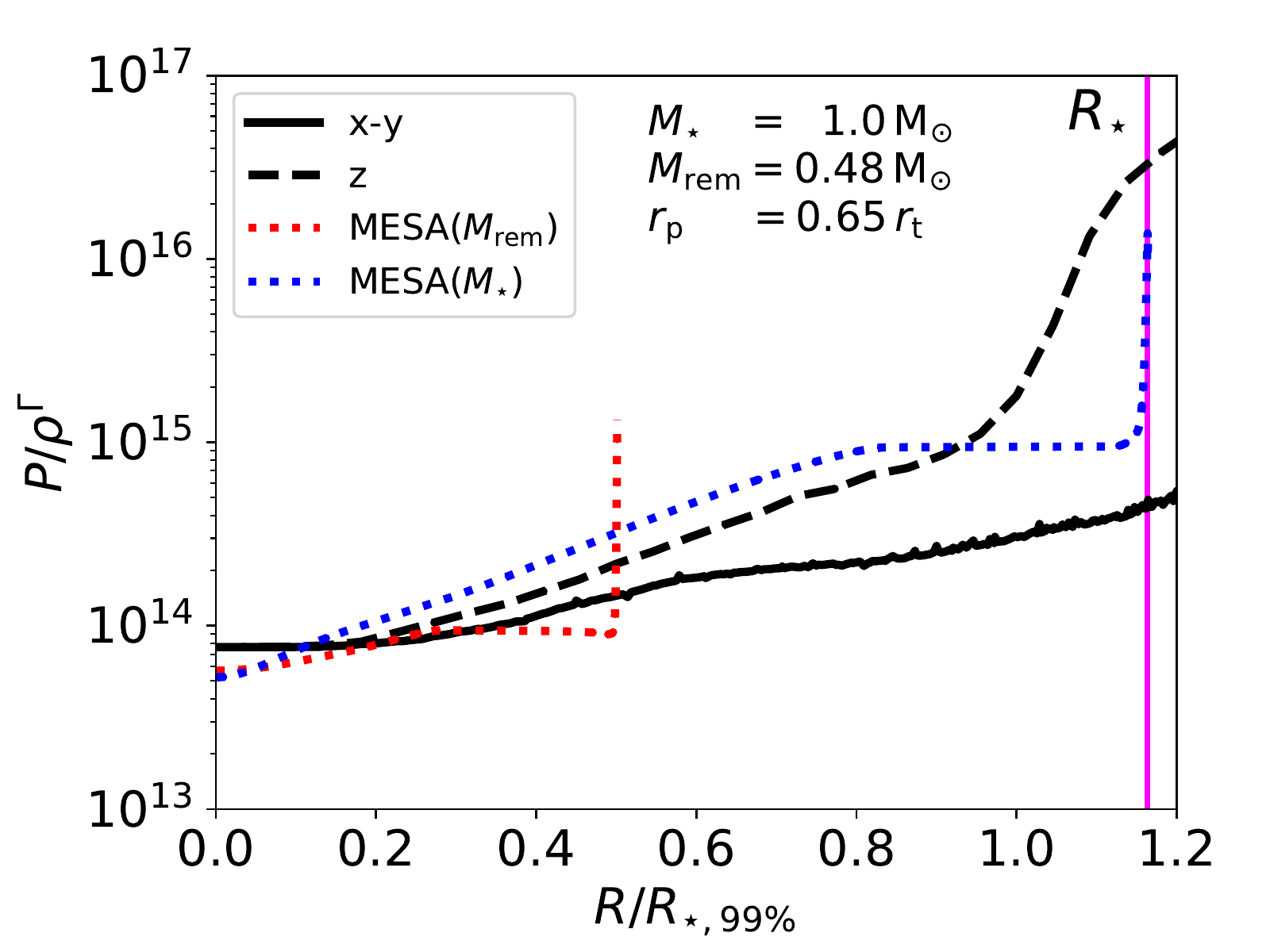}	\caption{ The specific entropy $P/\rho^{\Gamma}$ ($\Gamma=5/3$, in cgs units) of the same remnant star shown in Figure~\ref{fig:rotation} ($M_{\star}=1$, $M_{\rm rem}\simeq0.48$, $\psi=0.65$ and $r\simeq 23\rtidal$). The black curves represent the entropy profile in the equatorial plane ( azimuthally-averaged, solid) and along the $z-$axis (dashed).  The  blue and red dotted curves indicate the entropy profile for main sequence stars with mass $M_{\star}$ and $M_{\rm rem}$, respectively. The radius $R$ on the $x-$axis is normalized by the radius $R_{\star,99\%}$ for 99\% of the remnant mass. The vertical magenta solid line indicates the radius of the original  $1\Msol$ MS star, $R_{\star}$.
}
	\label{fig:entropy}
\end{figure}

\begin{figure}
	\centering
	\includegraphics[width=8.0cm]{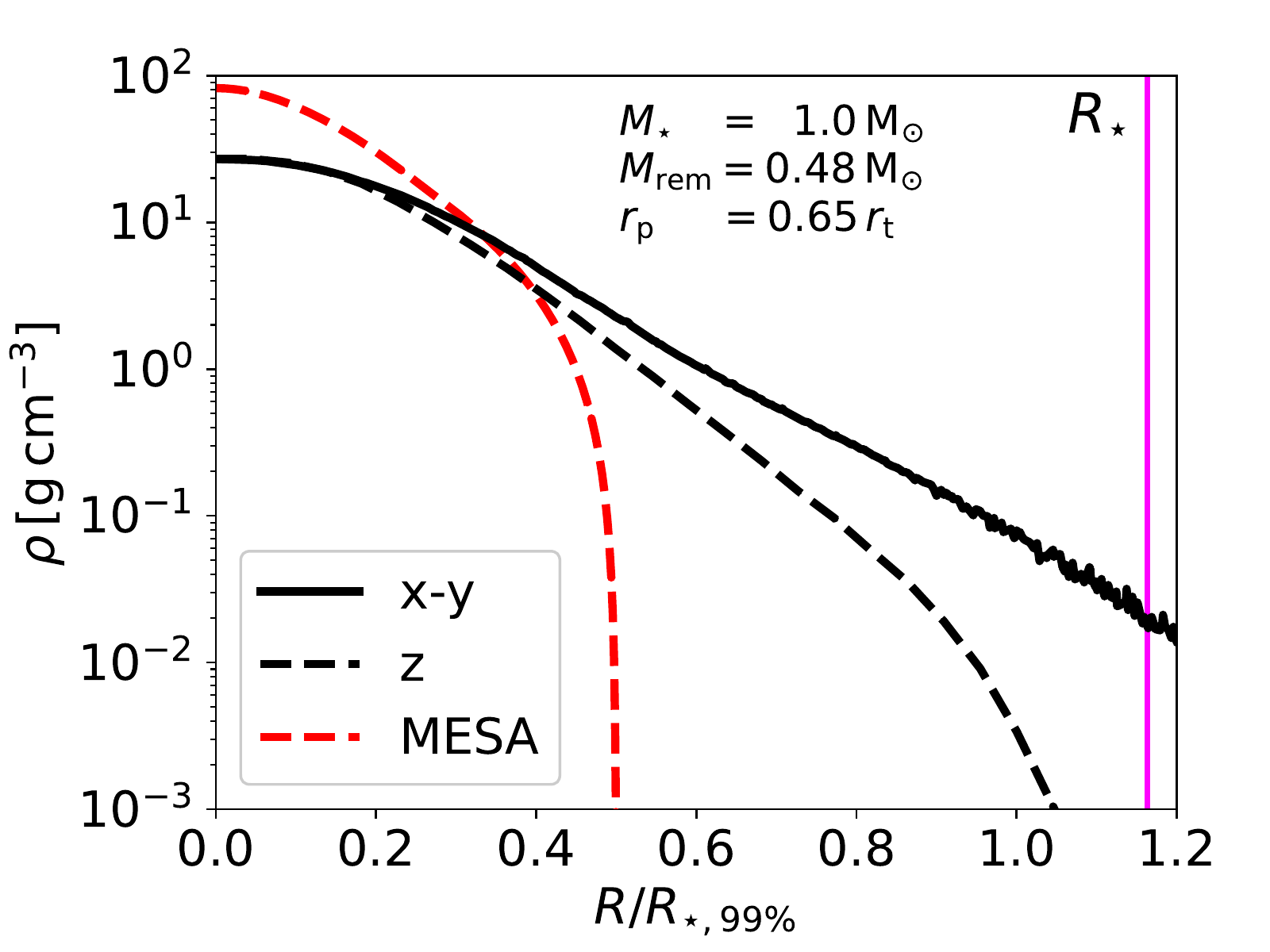}\vspace{-0.05in}
	\includegraphics[width=8.0cm]{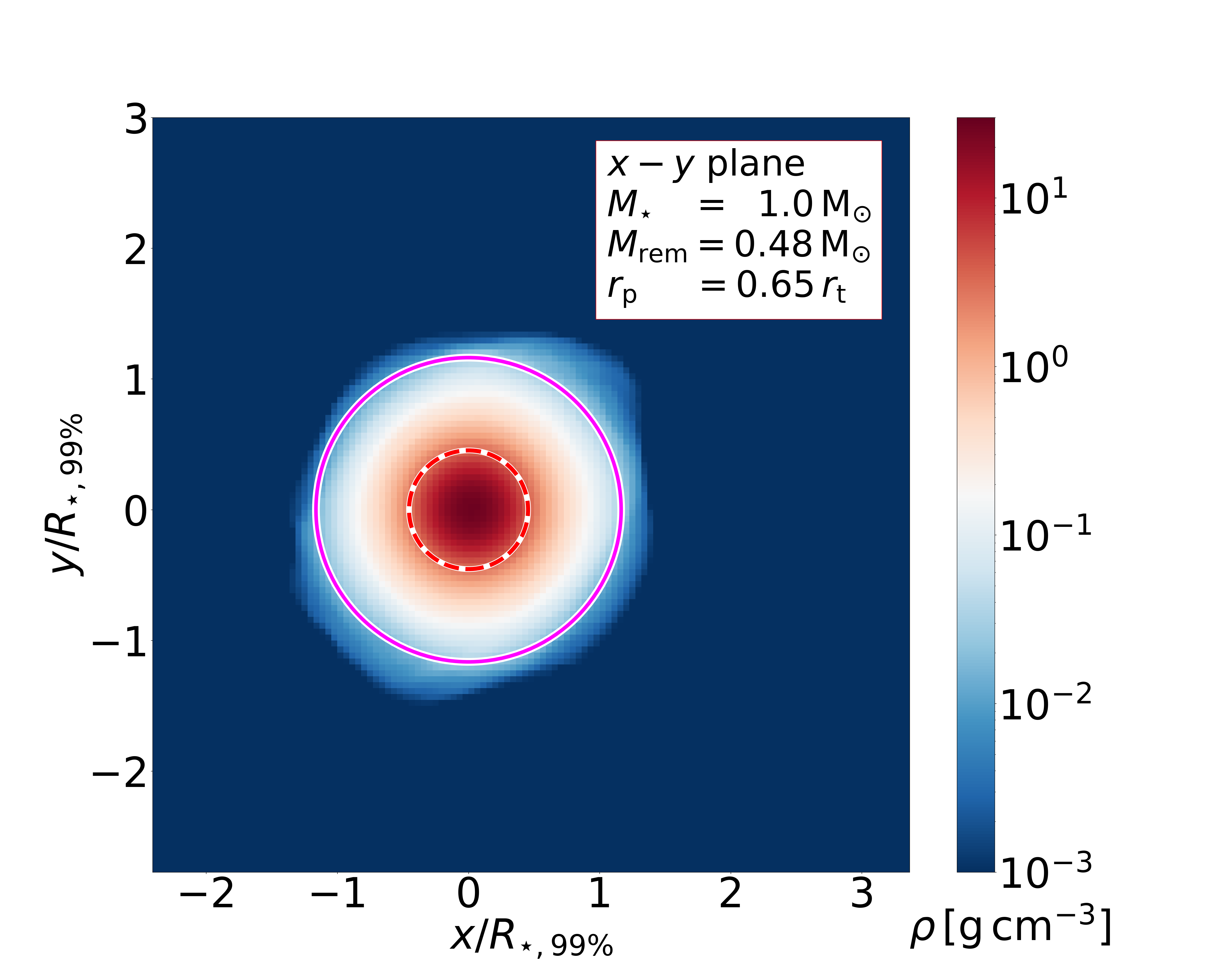}	\vspace{-0.1in}
	\includegraphics[width=8.0cm]{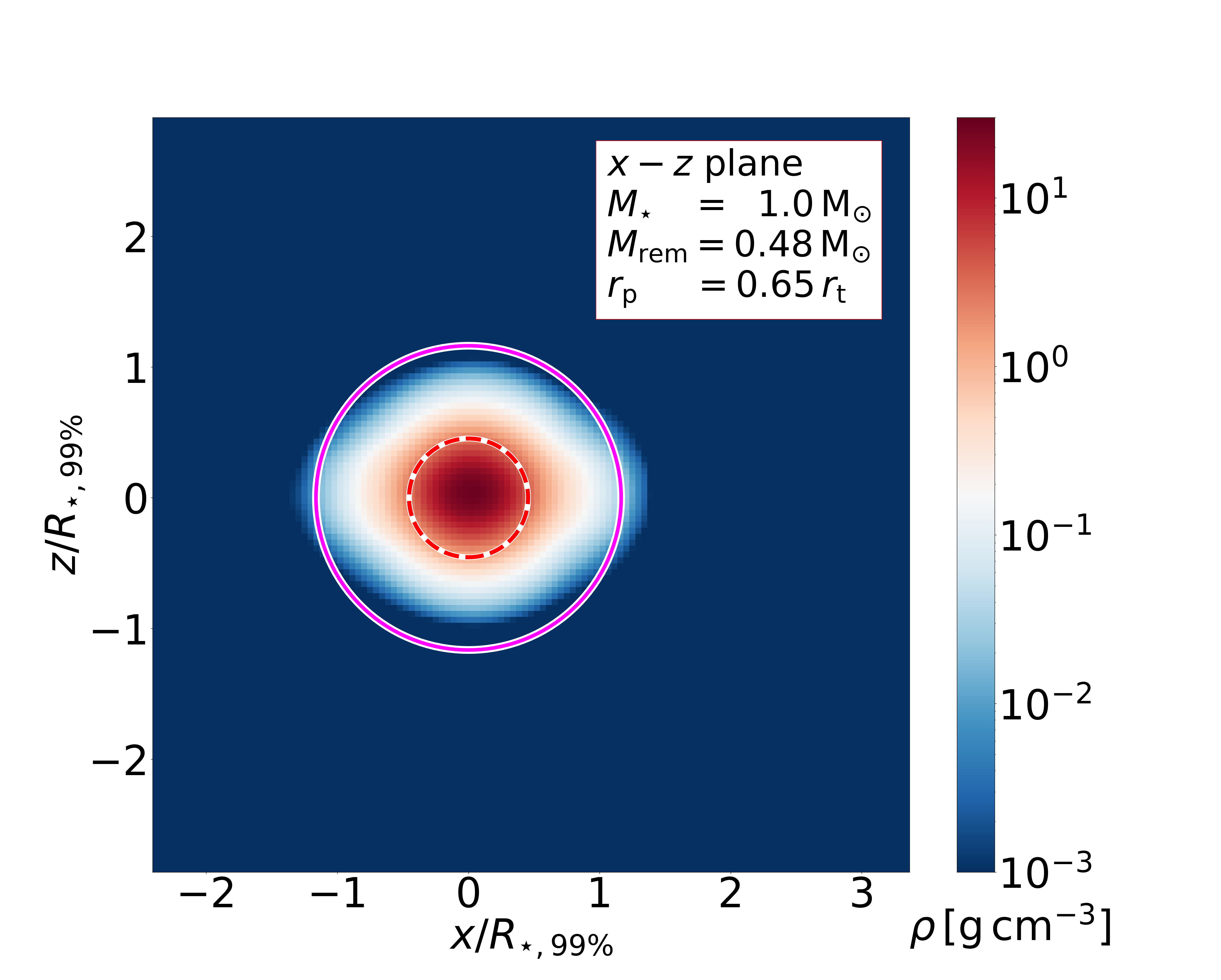}\\
	\caption{ The density $\rho$ of the same remnant star in Figure \ref{fig:rotation} ($M_{\star}=1$, $M_{\rm rem}\simeq0.48$, $\beta = 1.54$ and $r\simeq 23\rtidal$).
	The spatial scales ($R$, $x$, $y$) are normalized to $R_{\star,99\%}$, the radius containing 99\% of the remnant mass. (\textit{top} panel) The density on the equatorial plane is shown by the black solid curve and along the $z-$axis by the black dashed curve. The red dashed curve depicts the density profile of a \mesa-MS analog. The vertical magenta solid line indicates the original $R_{\star}$. (\textit{Middle} and \textit{bottom} panels) $2-$dimensional density maps of the star in the equatorial plane ($x-y$) and in the vertical plane ($x-z$), respectively. The solid (larger) magenta circle delineates $R_{\star}$ and the red (smaller) dashed line the radius for 99\% of the mass of the \mesa-MS star.}
	\label{fig:density}
\end{figure}

\subsubsection{Internal structure}
\label{subsub:density}

 Figure~\ref{fig:entropy} shows the specific entropy as a function of distance from the center of the star portrayed in Figure~\ref{fig:rotation}, a partial disruption of a $1 \Msol$ star that leaves a $0.48 \Msol$ remnant. 
 As we have assumed adiabatic behavior and found that the tidally-induced motions are laminar, the range of specific entropy found in the remnant matches the range found in the original star. However, the mean entropy in the remnant is a bit lower than in the initial star because $1\Msol$ main sequence stars have positive radial entropy gradients, and most of the mass lost in the encounter is taken from the star's outer layers.  Because the remnant rotates so rapidly, its specific entropy rises more gradually outward in the equatorial plane than along the rotational axis.

Although the specific entropy of the remnant is similar to that of the initial star, it is in general greater than in a main sequence star of the remnant mass because higher-mass main sequence stars have higher specific entropy than lower-mass stars.  For this reason, remnants of severe partial disruptions are, in general, far from thermal equilibrium.

A direct consequence of this departure from thermal equilibrium is shown in Figure~\ref{fig:density}, where we compare the density distributions of this remnant and its main-sequence counterpart.  The \textit{top} panel of Figure~\ref{fig:density} shows its density profile both in the equatorial plane and along the $z-$axis.  The density on the equatorial plane is calculated in the same way as the entropy in that plane. The \textit{middle} and \textit{bottom} panels depict $2-$dimensional snapshots of the star's density in the $x-y$ and $x-z$ planes, respectively.

The most noticeable feature in the top panel of Figure~\ref{fig:density} is how much more extended the density distribution is in the remnant than in its main-sequence counterpart.  At the center of the star, the density is about a factor of 3 smaller.  It is hard to determine an outer radius for this remnant because the density drops so smoothly outward: for radii outside $\simeq 0.3R_{\star}$, in the equatorial plane it is very well described by $\rho \propto \exp[-R/(0.15 R_{\star})]$.  The photosphere lies well outside the range portrayed: the Thomson optical depth over an exponential scale-length at $R/R_{\star} \simeq 1.2$ is $\sim 10^8$.  On the other hand, the majority of the star's mass is confined much more tightly, and can be found within a distance similar to that of a main sequence star of this mass, $\simeq 0.5R_{\star}$.

All these trends are reproduced in our other remnants, but, as might be expected, with the contrast between the remnant and its main sequence partner greater for more severe encounters.  In one case, the central density is a factor of 30 smaller than a main sequence star of the same mass.

All three panels of Figure~\ref{fig:density} portray
the star's oblate spheroidal shape. The top panel shows how the density drops outward more rapidly along the $z-$direction than on the equatorial plane.  The two lower panels show its shape in the equatorial and poloidal planes.  It is clear from them that, although the star is very nearly axisymmetric, it is substantially oblate, and the oblateness increases with distance from the center.

\section{Discussion - The fate  of the  stellar remnants}

\label{discuss:partialTD}

Due to slightly asymmetric mass-loss, remnants whose parent stars had very nearly zero energy with respect to the BH have a small, but non-zero, orbital energy per unit mass after their tidal encounters.   In real events, the initial stellar orbital energy can also be slightly non-zero, but the magnitude of the surviving  remnants' energy is sufficiently larger than the initial energy  that the latter  can be neglected (Section \ref{subsub:bound_unboud}).  The orbits of the remnants can then be conveniently divided into two classes according to the sign of their energy considering only the black hole potential: those with positive energy are unbound, and those with negative energy are bound.

\subsection{Unbound population}\label{discuss:unbound}

The ejection velocities of the unbound remnants we simulated range from $90-330\km\s^{-1}$. Extrapolating from the bulge dispersion data of galaxies with central BHs slightly more massive than $10^{6}\Msol$, we find that the dispersions of galaxy bulges containing BHs with  $M_{\rm BH}\simeq10^{6}$ are
$\sim \sigma=60-90\km/{\rm s}^{-1}$ \citep[e.g.,][]{Woo+2013, KormendyHo2013, Graham2016}. Our unbound remnants can therefore easily escape the radius of influence, $r_{\rm inf}= GM_{\rm BH}/\sigma^{2}\simeq 0.5\pc(M_{\rm BH}/10^{6})(\sigma/90\km\s^{-1})^{-2}$, of the central BH. Nonetheless, if the potential beyond the sphere of influence is logarithmic, the remnants are likely to reach a turning point $r_{\rm max}$ at only a few $r_{\rm inf}$, i.e., $r_{\rm max} \simeq \Lambda~r_{\rm inf}$ with $\Lambda = e^{(v_{\rm ejec}/2\sigma)^{2}}\simeq1-10$. Such a turning point would be well within the bulge region.

As the angular momentum of a remnant is much smaller than the value corresponding to a circular orbit at its semimajor axis, the pericenter distance is determined almost purely by the angular momentum.   If it is unchanged during the time spent near apocenter, any such remnant will return to the black hole with the same pericenter as the original stellar orbit, raising the prospect of a second tidal interaction.

To estimate how large these perturbations may be, as a crude approximation we compare the travel time for a partial disruption remnant to reach its turning point with the time required for weak stellar encounters to alter the remnant's original specific angular momentum $L_{0}$ by a factor of order unity.  The travel time is $t_{\rm travel}=r_{\rm max}/v_{\rm ejec}\simeq 10^{5} (r_{\rm  max}/10\pc)(v_{\rm ejec}/90\km\s^{-1})^{-1}\yr$.  On the other hand, the evolution time for remnant angular momentum is $t_{\rm L}\simeq (L_{0}/L_{\rm r})^{2}t_{\rm r}$ \citep{Merritt2013} where $L_{\rm r}$ refers to the specific angular momentum change during a collisional relaxation time $t_{\rm r}$; by definition, $L_{\rm r}\simeq\sigma~r_{\rm max}$. Using the relation $t_{\rm r}\simeq 0.1(N/\ln N)t_{\rm cross}$ \citep{BinneyTremaine1987}, where $N$ is the number of stars within the region the test-particle star travels through, the definitions $t_{\rm cross}=r_{\rm max}/\sigma$ and $r_{\rm max}=\Lambda~ r_{\rm inf}$, and the fact $L_{0}\simeq \sqrt{2GM_{\rm BH}r_{\rm t}}$, we find that the ratio between the two characteristic times when $r_{\rm max}>r_{\rm inf}$ is,
\begin{align}
\frac{t_{\rm L}}{t_{\rm travel}}&\simeq2 \left(\frac{0.1N}{\ln N}\right) \left(\frac{v_{\rm ejec}}{\sigma}\right)\left(\frac{ r_{\rm inf} r_{\rm t}}{r_{\rm max}^{2}}\right),\nonumber\\
&\simeq 10^{-2}\left(\frac{v_{\rm ejec}}{\sigma}\right)\left(\frac{M_{\rm BH}}{10^{6}}\right)^{4/3}\left(\frac{r_{\rm max}}{5\pc}\right)^{-1}.
\end{align}
For this estimate, we also assumed the mass of background stars is $1\Msol$, giving $N(<r_{\rm max})\simeq2~M_{\rm BH}~\Lambda$ for a logarithmic potential.

This estimate implies that gravitational encounters are very likely to result in changes of the unbound remnants' angular momenta large enough to alter their pericenter distances (a situation also called ``full loss-cone" evolution).  Because $r_{\rm p} \propto L^2$ for these highly-eccentric orbits, the  {resulting change}  of  $r_{\rm p}$ should be $\propto t_{\rm travel}/t_{\rm L}$. Thus, for these unbound remnants, the pericenter upon return is likely to be considerably larger than the value of $\rtidal$ of the returning remnant.  It is also possible for their angular momenta to be affected by other mechanisms, e.g. scattering by giant molecular clouds \citep{Perets+2007}  or torques due to non-spherical galactic pontentials \citep{MerrittPoon2004}. These remnants, although on unclosed  orbits, will nonetheless return to the galactic center close to the BH, but their pericenters are likely to be altered enough that the probability of an interesting tidal encounter is small.

\subsection{Bound population}
			\label{item:boundpopulation}
Every remnant in our bound sample (except for one that is exceptionally weakly bound) has an eccentricity less than unity by $\sim10^{-4}-10^{-5}$, a semimajor axis $a\sim0.03-0.5\pc$, and an orbital period $P\sim400-40000\yr$.  Although it is likely that our sample does not span the full range of possibilities, these numbers may be taken as indicative of the typical magnitudes for events with $M_{\rm BH} \sim 10^6$.

These bound remnants are also subject to stellar encounters, but within the black hole sphere of influence.
For this case, we can not use the same expression for $t_{\rm L}$ used above as it is derived for remnants whose motions are dominated by the potential from surrounding stars while, within $r_{\rm inf}$, the BH potential dominates.
The typical velocity of stars at $r_{\rm max}=2a<r_{\rm inf}$ is roughly $\sigma\simeq\sqrt{GM_{\rm BH}/2a}$. This leads to a relaxation time $t_{\rm r}=0.1 (M_{\rm BH}/m)^{2} /[N\ln(M_{\rm BH}/m)]~t_{\rm cross}$, where $m$ is the mean mass per star.
With $r_{\rm max}=2a$ and $t_{\rm travel}=P/2$, we find that $t_{\rm L}/t_{\rm travel}$ for our fiducial values is not very different from the value estimated for the unbound population:
\begin{align}
\label{eq:t_bound}
\frac{t_{\rm L}}{t_{\rm travel}}&\simeq \frac{0.1\times2^{3/2}}{\uppi }\frac{(M_{\rm BH}/m)^{2}}{N \ln (M_{\rm BH}/m)}\left(\frac{r_{\rm t}}{a}\right),\nonumber\\
&\simeq 2\times10^{-2} \left(\frac{N}{2\times10^{6}}\right)^{-1}\left(\frac{\ln (M_{\rm BH}/m)}{13.8}\right)^{-1},\nonumber\\
&\times m^{-2}\left(\frac{M_{\rm BH}}{10^{6}}\right)^{7/3}\left(\frac{a}{0.5\pc}\right)^{-1},
\end{align}
where we have scaled to values appropriate to the one of the longer semi-major axes in our sample. The apocenter distance for such a semi-major axis is comparable to $r_{\rm inf}$ for $M_{\rm BH}=10^{6}$, within which, by definition, $N(<r_{\rm inf})\simeq2\times10^{6}$. 

However, this timescale ratio is sensitive to the dependence of $N$ on $a$. If the stellar density $\rho_{\star}\propto r^{-n}$, $N(<r)\propto r^{3-n}$.  The ratio $t_{\rm L}/t_{\rm travel}$ then scales $\propto a^{n-4}$. Therefore, for a density profile near the BH with $n<4$, $t_{\rm L}/t_{\rm travel}$ increases as $a$ decreases, possibly becoming larger than unity at a sufficiently small $a$ (e.g., for $n=7/4$, the steady-state solution of \cite{BahcallWolf1976}, the ratio becomes larger than unity at $a\lesssim0.07-0.08\pc$). This means that for bound remnants with sufficiently small semimajor axes, the pericenter upon return remains almost unchanged from its value during the first passage. Because our sample includes some remnants with semimajor axes as small as $\simeq 0.03$~pc, a fraction of the bound remnant population will return with pericenters either the same as during their first passage, or enlarged by only a little. 

\subsection{A second tidal disruption?}

Whether a significant tidal disruption event takes place at the next pericenter passage depends on how the (possibly larger) pericenter compares to the star's new tidal radius.  If the remnant returns to the main sequence before returning to the vicinity of the black hole, its smaller mass would imply a smaller size and a smaller $r_{\rm t}$, whereas its new pericenter is likely to be at least as large as in the original event.  Significant disruption would probably not occur.

However, return to the main sequence in time for the next return to pericenter may be problematic.  Relative to main sequence structure, these remnants are expanded by both extra heat and rapid rotation.  In terms of its enclosed mass profile, the example shown in Figure~\ref{fig:density} resembles a red giant: most of its mass is contained within a relatively small radius, while a low-density envelope extends out to large distances.  Employing our semi-analytic model \citepalias{Ryu1+2019}, we might then estimate a critical distance for complete disruption $\simeq 1.5 \times$ that expected for the same-mass main sequence star, which is $\simeq 1.8~\physrad$ for the parent star, but a critical distance for partial disruptions $\simeq 1.4\times$ that of the parent star.  Both distances are also enlarged by a modest amount because the ratio $(M_{\rm BH}/M_{\star})^{1/3}$ is greater by 28\%.  Thus, if there is too little time for it to cool before the next pericenter passage, a significant tidal encounter might well take place upon its first return to the vicinity of the black hole.

Whether thermal relaxation can be completed by the time the remnant returns to periastron depends upon the ratio of the cooling time to the orbital period.
The photon diffusion time from the center of a star to its edge is
		\begin{align}
t_{\rm th}&\simeq \kappa_{\rm c}\rho_{\rm c} R_{\rm c}^{2} /c,\nonumber\\
&\simeq 2\times10^{4}\yr\left(\frac{\kappa_{\rm c}}{10\kappa_{\rm e}}\right)\left(\frac{\rho_{\rm c}}{10^{2}\gram\cm^{-3}}\right)\left(\frac{R_{\rm c}}{0.1}\right)^{2},
\end{align}
where $\kappa_{\rm c}$ is the core opacity, $\rho_{\rm c}$ is the core density and $R_{\rm c}$ is the radial length scale of the core. The Thomson opacity is $\kappa_{\rm e}$.  In the conditions of our stellar remnants ($\rho_{\rm c}\sim 1- 10^{2} \gram\cm^{-3}$, core temperature $T_{\rm c}\sim 10^{6} - 10^{7} \K$), $\kappa_{\rm c}/\kappa_{\rm e}\sim1- 10^{2}$ \citep{Hayashi+1962}. Comparing this time to the orbital periods shown in Table~\ref{tab:boundunbound} demonstrates that the more tightly bound remnants ($P<t_{\rm th}$) would return back to the BH without significant changes in their internal structures. These are also the remnants likely to suffer the least increase in orbital pericenter due to scattering with background stars.  Thus, for both reasons, the more tightly bound remnants have the greatest probability of going through a second TDE.

However, we caution that a more careful calculation of the remnant's cooling is necessary to determine what happens when it next passes through pericenter. The evolution of the remnant star's rotation may also influence its fate. Angular momentum may be lost through magnetic braking \citep[e.g][]{FrickeKippenhahn1972}; it may also be mixed inward from the outer $\sim10\%$ of the star's mass where it initially resides by any of a variety of processes \citep{MaederMeynet2000}.  Because only a minority of the remnants' mass rotates rapidly, evolution in the star's rotation may be a next-order correction to the effect of cooling.

\section{Summary}
\label{sec:summary}

In this paper, the third in this series, we continue our study of tidal disruption events of main-sequence stars, focusing on the properties of partial disruptions. Our results are based upon a suite of fully general relativistic simulations in which the stars' initial states are described by realistic main-sequence models. We examined tidal disruption  events for eight different stellar masses, from $M_{\star}=0.15$ to $M_{\star}=10$ with a fixed black hole mass ($10^6\Msol$).
In \citetalias{Ryu4+2019}, we will explore how increasingly strong relativistic effects alter the properties of partial disruptions involving higher-mass black holes.

We find that the energy distribution $dM/dE$ of the stellar debris created from partial disruptions is different from the one that arises in  full disruptions, with the contrast growing for weaker encounters. For full disruptions, the characteristic energy width $\Delta E$ of the stellar debris for low-mass stars is $\simeq 0.8 \Delta\epsilon$, while that for high-mass stars can be as large as $\simeq 2\Delta\epsilon$, where $\Delta\epsilon$ is the traditional order of magnitude estimate for this width.  The energy distribution $dM/dE$ for all masses has a local minimum near $E\simeq0$ and ``shoulders'' near the outer boundaries, with a contrast between the two $\simeq 1.5$ (\citetalias{Ryu2+2019}). On the other hand, for partial disruptions, most of the mass of the stellar debris is concentrated near the shoulders, with little mass near $E\simeq 0$: the contrast is $\sim 10$ for strong disruptions, in which a large fraction of the stellar mass is lost, and it increases to $\sim 100-1000$ for weaker disruptions. Although the outer edges of the distribution are quite sharp for low-mass stars subjected to either partial or full disruption, there can be significant tails for high-mass stars.  These become progressively steeper for weaker partial disruptions. Because there is so little mass near $E\simeq0$, late-time fallback is suppressed, and the overall shape of the fallback rate becomes more and more like a single peak as the mass lost in the event diminishes.  On the declining side of the peak, the mass-return rate is $\propto t^{-p}$ with $p\simeq2-5$, very unlike the consistent $p=5/3$ for full disruptions. 
	
Another product of partial disruptions is surviving remnants.
We have found a simple analytic expression linking the ratio between the stellar orbit's pericenter and the physical tidal radius for that stellar mass to the ratio between the remnant mass and the original stellar mass (see Equation~\ref{eq:bestfit_remmass1}).
The remnants retain around $50\%$ of the original mass at $r_{\rm p}/\mathcal{R}_{\rm t}\simeq1.2-1.5$, while the mass loss becomes less than $10\%$ at $r_{\rm p}/\mathcal{R}_{\rm t}\gtrsim1.5-1.8$. 

Because higher-mass main sequence stars have higher entropy than lower-mass stars, surviving remnants are out of thermal equilibrium and tend to be larger in size than a MS star of the same mass.  They are also rapidly-rotating, reaching angular frequencies near break-up in the outer layers of the remnants left by events causing substantial mass-loss from initially massive stars. The rapid rotation makes these stars oblate spheroids. 

The change in specific orbital energy of partially-disrupted stars is quite small compared to the spread in energy of the debris: $\simeq10^{-3} \Delta \epsilon$ (see Table~\ref{tab:boundunbound}), but it can be of either sign.  Particularly for low-mass stars, weaker encounters lead to remnants that lose orbital energy and therefore remain within the sphere of influence of the black hole, while the strongest encounters can create remnants able to travel some distance out into the galaxy's bulge.  For high-mass stars, most partial disruptions lead to bound remnants, except for those that are nearly strong enough to cause total disruption.

When a stellar remnant, whether bound to the black hole or able to travel out into the bulge, reaches its orbital apocenter, weak gravitational interactions with buldge  stars can alter its angular momentum.  The change can be large compared to the remnant's original angular momentum when the remnant goes as far as the stellar bulge, or even the outer portion of the black hole's sphere of influence, but if the remnant's apocenter is smaller than the black hole's sphere of influence, the change can be comparable to the original angular momentum or even less.  When the increase in specific angular momentum is relatively small, the remnant may become a victim  of another TDE if its cooling time is longer than its orbital period.  Because the most tightly-bound remnants have substantially shorter orbital periods than those able to reach the bulge, their prospects for a second tidal event are further enhanced.

\section*{Acknowledgements}

 We would like to thank an anonymous referee for an insightful question about the specific entropy in partial disruption remnants. This work was partially supported by NSF grant AST-1715032, Simons Foundation grant 559794 {and an advanced ERC grant TReX}. S.~C.~N. was supported by the grants NSF AST 1515982, NSF OAC 1515969, and NASA 17-TCAN17-0018, and an appointment to the NASA Postdoctoral Program at the Goddard Space Flight Center administrated by USRA through a contract with NASA.  This research project (or part of this research project) was conducted using computational resources (and/or scientific computing services) at the Maryland Advanced Research Computing Center (MARCC). The authors would like to thank Stony Brook Research Computing and Cyberinfrastructure, and the Institute for Advanced Computational
Science at Stony Brook University for access to the high-performance
SeaWulf computing system, which was made possible by a $\$1.4$M National Science Foundation grant (\#1531492).

\software{
matplotlib \citep{Hunter:2007}; \mesa \citep{Paxton+2011}; 
\harm  \citep{Noble+2009};
}

\end{document}